\documentclass[conference]{IEEEtran}
\IEEEoverridecommandlockouts
\usepackage{cite}
\usepackage{amsmath,amssymb,amsfonts}
\usepackage{algcompatible}
\usepackage{algpseudocode}
\usepackage[linesnumbered,ruled,vlined]{algorithm2e}
\usepackage{graphicx}
\usepackage{listings}
\usepackage{textcomp}
\def\BibTeX{{\rm B\kern-.05em{\sc i\kern-.025em b}\kern-.08em
    T\kern-.1667em\lower.7ex\hbox{E}\kern-.125emX}}
\usepackage{pifont}

\newcommand{\name}{\textsc{BlockGPT}\xspace}

\DeclareMathOperator*{\argmax}{arg\,max}
\DeclareMathOperator*{\argmin}{arg\,min}

\newcommand{\TotalNumTxs}{\empirical{$68M$}\xspace}
\newcommand{\DataStartBlock}{\block{$5,470,817$}\xspace}
\newcommand{\DataStartDate}{\empirical{$19$th~April,~$2018$}\xspace}
\newcommand{\DataEndBlock}{\block{$15,000,000$}\xspace}
\newcommand{\DataEndDate}{\empirical{$21$st~June,~$2022$}\xspace}
\newcommand{\DataNumOfDays}{\empirical{$1,523$}\xspace}
\newcommand{\CaseOneNumOfAttacks}{\empirical{$124$}\xspace}

\newcommand{\cid}{C_\text{ID}}

    \usepackage[table]{xcolor}

\usepackage{algorithm2e}
\RestyleAlgo{ruled} 
\SetAlFnt{\small}
\usepackage{amsmath}
\usepackage{amssymb} 
\usepackage{amsthm}

\usepackage{booktabs} 
\usepackage{cancel}
\usepackage{enumitem}
\setlist[itemize]{leftmargin=*,itemsep=-4pt,noitemsep}
\setlist[enumerate]{leftmargin=*,itemsep=-4pt,noitemsep}
\setlist[description]{leftmargin=*,itemsep=-4pt,noitemsep}
\usepackage{graphicx}
\usepackage{makecell}
\usepackage{pifont} 
\usepackage{hyperref}
\usepackage{xspace}
\usepackage{multirow}

\newcommand{\circled}[1]{\raisebox{.5pt}{\textcircled{\raisebox{-.9pt} {{#1}}}}}

\newcommand{\empirical}[1]{#1}

\usepackage{xstring}
\newcommand{\etal}{\textit{et al.\ }}
 
\newcommand{\block}[1]{\href{https://etherscan.io/block/#1}{#1}\xspace}

\newcommand{\abbrEtherscanTx}[1]{\href{https://etherscan.io/tx/#1}{\StrLeft{#1}{6}..\StrRight{#1}{4}}\xspace}

\newcommand{\abbrEtherscanAddress}[1]{\href{https://etherscan.io/address/#1}{\StrLeft{#1}{6}..\StrRight{#1}{4}}\xspace}

\definecolor{gainsboro}{rgb}{0.86, 0.86, 0.86}
\newcommand{\clg}[1]{\cellcolor{gainsboro}{#1}}

\usepackage{tikz}
\newcommand*\ec[1][1ex]{\tikz\draw (0,0) circle (#1);} 
\newcommand*\hc[1][1ex]{%
  \begin{tikzpicture}
  \draw[fill] (0,0)-- (90:#1) arc (90:270:#1) -- cycle ;
  \draw (0,0) circle (#1);
  \end{tikzpicture}}
\newcommand*\fc[1][1ex]{\tikz\fill (0,0) circle (#1);} 
\usepackage{acro}
\acsetup{single}

\DeclareAcronym{ABI}{
  short = ABI,
  long  = Application Binary Interface,
}
\newcommand{\ABI}{\ac{ABI}\xspace}

\DeclareAcronym{AMM}{
  short = AMM,
  long  = Automated Market Maker,
}

\DeclareAcronym{API}{
  short = API,
  long  = Application Programming Interface,
}

\DeclareAcronym{BEV}{
  short = BEV,
  long  = Blockchain Extractable Value,
}
\newcommand{\BEV}{\ac{BEV}\xspace}

\DeclareAcronym{BRF}{
  short = BRF,
  long  = Back-run Flodding,
}

\DeclareAcronym{CeFi}{
  short = CeFi,
  long  = Centralized Finance,
}

\DeclareAcronym{DApp}{
  short = DApp,
  long  = Decentralized Application,
}
\newcommand{\DApp}{\ac{DApp}\xspace}
\newcommand{\DApps}{\acp{DApp}\xspace}

\DeclareAcronym{DeFi}{
  short = DeFi,
  long  = Decentralized Finance,
}
\newcommand{\DeFi}{\ac{DeFi}\xspace}

\DeclareAcronym{DEX}{
  short = DEX,
  long  = Decentralized Exchange,
}

\DeclareAcronym{DNN}{
  short = DNN,
  long  = Deep Neural Network,
}

\DeclareAcronym{DOS}{
  short = DOS,
  long  = Denial-of-Service,
}
\newcommand{\DOS}{\ac{DOS}\xspace}

\DeclareAcronym{EV}{
  short = EV,
  long  = Extractable Value,
}

\DeclareAcronym{EVM}{
  short = EVM,
  long  = Ethereum Virtual Machine,
}
\newcommand{\EVM}{\ac{EVM}\xspace}

\DeclareAcronym{FaaS}{
  short = FaaS,
  long  = Front-running as a Service,
}
\newcommand{\FaaS}{\ac{FaaS}\xspace}

\DeclareAcronym{IDS}{
  short = IDS,
  long  = Intrusion Detection System,
}
\newcommand{\IDS}{\ac{IDS}\xspace}

\DeclareAcronym{ITR}{
  short = ITR,
  long  = Intermediate Trace Representation,
}
\newcommand{\ITR}{\ac{ITR}\xspace}

\DeclareAcronym{LSTM}{
  short = LSTM,
  long  = Long Short-Term Memory,
}

\DeclareAcronym{MEV}{
  short = MEV,
  long  = Miner Extractable Value,
}

\DeclareAcronym{NLP}{
  short = NLP,
  long  = Natural Language Processing,
}
\newcommand{\NLP}{\ac{NLP}\xspace}

\DeclareAcronym{P2P}{
  short = P2P,
  long  = peer-to-peer,
}
\newcommand{\PtP}{\ac{P2P}\xspace}

\DeclareAcronym{PGA}{
  short = PGA,
  long  = Priority Gas Auction,
}

\DeclareAcronym{PoW}{
  short = PoW,
  long  = Proof of Work,
}
\newcommand{\PoW}{\ac{PoW}\xspace}

\DeclareAcronym{PoS}{
  short = PoS,
  long  = Proof of Stake,
}
\newcommand{\PoS}{\ac{PoS}\xspace}

\DeclareAcronym{RNN}{
  short = RNN,
  long  = Recurrent Neural Network,
}

\DeclareAcronym{SOTA}{
  short = SOTA,
  long  = State-of-the-Art,
}
\newcommand{\SOTA}{\ac{SOTA}\xspace}

\DeclareAcronym{TVL}{
  short = TVL,
  long  = Total Value Locked,
}
\newcommand{\TVL}{\ac{TVL}\xspace}

\DeclareAcronym{IPS}{
  short = IPS,
  long  = Intrusion Prevention System,
}
\newcommand{\IPS}{\ac{IPS}\xspace}

\DeclareAcronym{FPR}{
  short = FPR,
  long  = False Positive Rate,
}
\newcommand{\FPR}{\ac{FPR}\xspace}

\begin{document}

\title{Blockchain Large Language Models}

\author{
\IEEEauthorblockN{
Yu Gai\IEEEauthorrefmark{1}  \IEEEauthorrefmark{2}\IEEEauthorrefmark{5}, 
Liyi Zhou\IEEEauthorrefmark{1}  \IEEEauthorrefmark{3}\IEEEauthorrefmark{5}
Kaihua Qin  \IEEEauthorrefmark{3}\IEEEauthorrefmark{5}, 
Dawn Song  \IEEEauthorrefmark{2}\IEEEauthorrefmark{5},
and 
Arthur Gervais  \IEEEauthorrefmark{4}\IEEEauthorrefmark{5}
}

\IEEEauthorblockA{
\IEEEauthorrefmark{2}University of California, Berkeley,
\IEEEauthorrefmark{3}Imperial College London,
\IEEEauthorrefmark{4}University College London,
\\
\IEEEauthorrefmark{5}Berkeley Center for Responsible, Decentralized Intelligence (RDI)
\\
}
}

\maketitle

\begingroup\renewcommand\thefootnote{\IEEEauthorrefmark{1}}
\footnotetext{Both authors contributed equally to this work.}

\begin{abstract}
This paper presents a dynamic, real-time approach to detecting anomalous blockchain transactions. The proposed tool, \name, generates tracing representations of blockchain activity and trains from scratch a large language model to act as a real-time Intrusion Detection System. Unlike traditional methods, \name is designed to offer an unrestricted search space and does not rely on predefined rules or patterns, enabling it to detect a broader range of anomalies. We demonstrate the effectiveness of \name through its use as an anomaly detection tool for Ethereum transactions. In our experiments, it effectively identifies abnormal transactions among a dataset of $68M$ transactions and has a batched throughput of $2284$ transactions per second on average. Our results show that, \name identifies abnormal transactions by ranking $49$ out of $124$ attacks among the top-$3$ most abnormal transactions interacting with their victim contracts. This work makes contributions to the field of blockchain transaction analysis by introducing a custom data encoding compatible with the transformer architecture, a domain-specific tokenization technique, and a tree encoding method specifically crafted for the \EVM trace representation.
\end{abstract}

\section{Introduction}\label{sec:introduction}
With the increasing number of transactions per second processed by blockchains, a rich real-world dataset of user behavior and \DApp interactions has become accessible across the globe. For the first time in history, the information security community can access a transparent, timestamped, and non-repudiable dataset of transactions, including their dynamic smart contract execution traces. This dataset also contains attack transactions that have caused multi-million-dollar losses. Between April 30, 2018, and April 30, 2022, users, liquidity providers, speculators, and protocol operators in blockchain networks suffered a total loss of at least $3.24$ billion USD~\cite{zhou2022sok}. These significant losses underscore the need for more generic, dynamic and scalable approaches to detect anomalous blockchain transactions, especially as the volume of transaction data continues to grow. 

Real-time \IDS for blockchain transactions, however, remains challenging due to constraints on the search space and the substantial manual engineering efforts required. More specifically, \SOTA works predominantly employ either \textit{(i)} reward-based approaches, which focus on identifying and exploiting transactions that yield significant profits, or \textit{(ii)} pattern-based techniques that depend on custom heuristics to deduce blockchain transaction intents and user address behavior (cf.~Figure~\ref{tab:comparison}). However, the reliance on predefined rules, patterns, or profitable vulnerabilities may prevent these methods from capturing the full spectrum of anomalies. Consequently, there is an urgent need for more versatile and adaptive techniques that can effectively identify a wide range of anomalous transactions in real-time, enhancing the security of blockchain networks.

When exploring the different means to perform anomaly detection, it becomes apparent that a blockchain transactions' trace outlines what a transaction does: the invoked smart contracts, the corresponding parameters, the order of invocations and storage information. Given that an attack transaction often executes a different path and exhibits different execution behaviors than normal transactions, we conjecture that an attack transaction portrays a different trace representation than benign transactions. Thus, we hypothesize that it is possible to train a model to learn representations of transaction execution traces without any prior knowledge of vulnerability patterns (e.g., reentrancy, price oracle manipulation, etc.) or information on transaction profitability.  Certainly, there may also be limitations to such an approach, such as the number of false positives~---~we will also explore the efficacy of this approach with suitable detection or anomaly thresholds.

\begin{figure}[tb]
    \centering
    \includegraphics[width=\columnwidth]{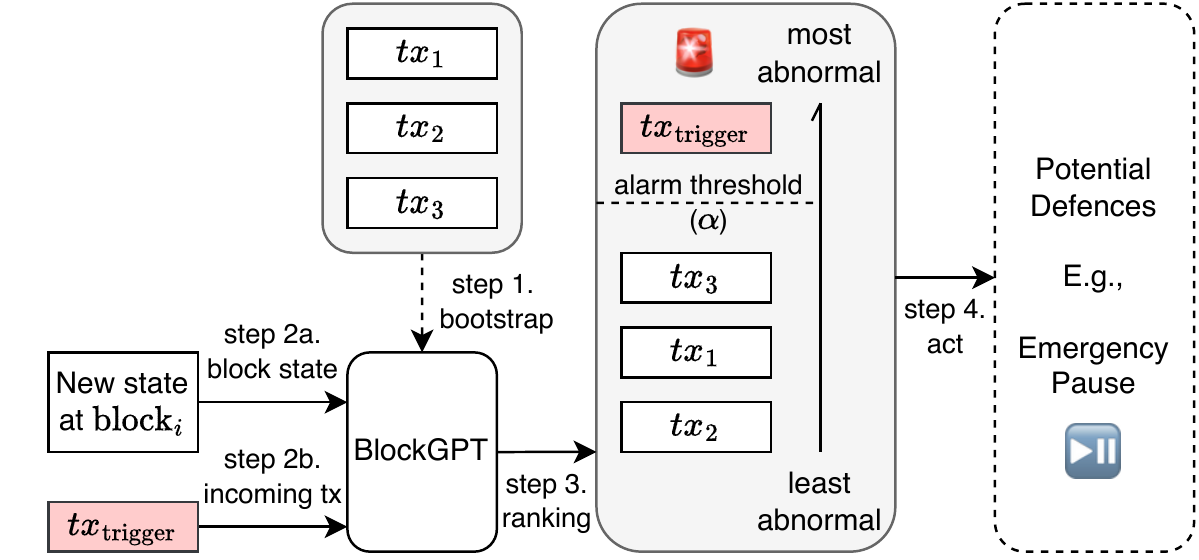}
    \caption{High-level overview of the \name defense mechanism, which consists of the following four major steps. \ding{182}~\name is bootstrapped by feeding in a dataset of historical transactions to train the model using unsupervised learning. \ding{183}~Depending on the system and threat model, \name detects new block states, including already confirmed transactions, and pending transactions (cf. Section~\ref{sec:models}). \ding{184}~\name ranks transactions based on how abnormal their execution traces are (cf. Section~\ref{sec:details}). \ding{185}~If an abnormal transaction is detected (cf. Section~\ref{sec:evaluation}), \name triggers a defense mechanism such as a front-running emergency pause.}
    \label{fig:highleveloverview}
\end{figure}

In this work, we leverage Ethereum as an open dataset to train a transaction anomaly ranking tool, \name. The main objective of \name is to elevate the art of blockchain security analysis into a comprehensive and real-time endeavor. By generating efficient trace representations of blockchain activity, \name trains a scalable large language model from scratch using an extended and augmented dataset from a previous study, covering attacks within a four-year timeframe. This entire dataset comprises a total of \TotalNumTxs transactions, encompassing all transactions from \CaseOneNumOfAttacks smart contracts that experienced an attack during the specified period. We evaluate the effectiveness of \name as a transaction anomaly ranking tool on the same dataset, assessing its performance in identifying anomalous transactions.

We evaluate \name using two simple metrics: \textit{(i)} a \emph{percentage ranking alarm threshold}; and \textit{(ii)} an \emph{absolute ranking alarm threshold}. \name successfully identifies abnormal transactions, ranking $49$ out of the $124$ attacks among the top-3 most abnormal transactions interacting with their corresponding victim contracts. Specifically, \name detects 20 adversarial transactions as the most abnormal, 20 as the second most abnormal, and 7 as the third most abnormal transaction associated with their victim contracts.

Regarding the practicality of \name, our analysis highlights its effectiveness when dealing with many transactions. For example, in DeFi applications with over $10,000$ transactions and a $0.01\%$ alarm threshold, \name successfully detects $24\%$ of the attacks and maintains an average \FPR of $0.097\%$ (cf. Figure~\ref{fig:threshold}). In the context of popular \DeFi applications processing $100$ transactions per day, a $0.1\%$ \FPR generates one alert approximately every 10 days. This demonstrates that \name is capable at providing a manageable number of alerts for further investigation, making it particularly suitable for high-volume transaction environments.

In terms of performance, \name showcases its real-time capabilities, achieving an average batched throughput of $2,284 \pm 289$ transactions per second, and on average $0.16 \pm 0.3$ seconds to rank a single transaction. This rapid detection of malicious blockchain transactions enables the triggering of a smart contract pause mechanism to prevent an attack as an \IPS. Approximately $50\%$ of the attacked contracts we investigate already have such a pause mechanism deployed. 

In summary, this work makes the following contributions to the field of blockchain transaction analysis:

\begin{itemize}
    \item To our knowledge, this paper is the first to apply unsupervised/self-supervised learning for anomaly detection in smart contract transaction execution traces. We develop a large language model for Ethereum transaction anomaly detection, employing custom data encoding, domain-specific tokenization, and a tree encoding method tailored for \EVM trace tree representation, capturing calls, function names, parameters, and storage modifications.

    \item We apply \name as an anomaly detection tool for Ethereum transactions to identify suspicious or malicious activities on the blockchain. We evaluate \name on a dataset of~\CaseOneNumOfAttacks attacks, consisting of a total of~\TotalNumTxs transactions, spanning a period of~\DataNumOfDays days, starting from block~\DataStartBlock (\DataStartDate) and ending at block~\DataEndBlock (\DataEndDate). We analyze the model across various dataset sizes and metrics, including two alarm thresholds, F1-score, F10-score, and CID score, demonstrating \name's robustness and versatility. We benchmark \name against doc2vec and trace length heuristics, showcasing the superior effectiveness of~\name.    

    \item Evaluation results indicate \name effectively identifies abnormal transactions and can detect different types of malicious activities, as shown through a flash loan attack case study. With a throughput of $2284 \pm 289$ transactions per second, our tool is a viable real-time \IDS for blockchains.

\end{itemize}

\section{Background}\label{sec:background}

\subsection{Blockchain and \DeFi}

\subsubsection{Blockchain} Since the inception of blockchains with Bitcoin in~$2008$~\cite{bitcoin}, it became apparent that their most well-suited use case is the transfer or trade of financial assets without trusted intermediaries~\cite{wust2018you}. A blockchain is considered permissionless when entities can join and leave the network at any time. Users authorize transactions through a public key signature and a subsequent broadcast on the blockchain \PtP network. Due to the openness of the \PtP network, the information about a transaction becomes public, once a transaction is broadcast. For example, blockchain participants can observe which smart contract a pending transaction calls triggers along with the corresponding call parameters. Miners accumulate unconfirmed transactions and solve a \PoW puzzle to append blocks to the blockchain. Various alternatives to \PoW, such as \PoS~\cite{saleh2021blockchain,bano2019sok} emerged. In addition to the block reward and transaction fees, \BEV is a new miner reward source~\cite{qin2022quantifying}. For a more thorough blockchain background, we refer the reader to SoKs~\cite{bonneau2015sok,atzei2017survey,bano2017consensus}.

\subsubsection{Smart Contracts}
While Bitcoin supports basic smart contracts through a stack-based scripting language, the addition of support for higher-level programming languages (e.g., Solidity) has resulted in widespread adoption.
Note that \SOTA blockchains generally require transaction fees as in to prevent \DOS attacks. 
Smart contracts are therefore only quasi Turing-complete because their execution can suddenly interrupt if the transaction fees exceed a predefined amount.
Notably, blockchains do not store the human-readable source code, nor the application interface to interact with a smart contract (i.e., \ABI). Instead, \SOTA blockchains only store the compiled bytecode on-chain.

The execution of a smart contract's bytecode is triggered by blockchain transactions, which are then carried out within so-called virtual machines (e.g., \EVM).
Similar to traditional programming languages such as Java, the execution of smart contracts can be summarized with execution traces (also known as logs), which record the state transition at each step of the process.
While full blockchain clients store the entirety of the historical blocks, intermittent states and execution traces are typically discarded due to excessive storage requirements. So-called archive nodes, however, store and provide an indexed database of historical smart contract executions.

\subsubsection{\DeFi} \DeFi refers to an ecosystem of financial products and services built on top of permissionless blockchains. \DeFi is currently experiencing a surge in popularity, with a peak \TVL \href{https://defillama.com/}{reaching $250B$~USD in December~$2021$}. Despite incorporating basic functions inspired by traditional finance (e.g., lending, trading, derivatives and asset management), \DeFi also introduces more novel designs (e.g., flash loans~\cite{qin2021attacking}, automated market makers~\cite{zhou2021high} and composable trading~\cite{zhou2021just}) enabled by a blockchain's atomicity property and \DeFi's composable nature.
Understanding the semantics of transactions that trigger these novel \DeFi designs presents a particular challenge because \DeFi transactions typically are intertwined with multiple financial \DApps.

\subsection{\NLP}
Natural language models are designed to process and generate human-like text. They are used in a wide range of applications, including language translation, text summarization, language generation, and text classification.

\subsubsection{\NLP and Bytecode}
There have been several approaches to applying natural language models to code, assembly, and bytecode. One approach is to treat the code or assembly as natural language text and input it into a natural language model. This can be useful for tasks such as code summarization, code generation, or code translation.

Another approach is to first convert the code or assembly into a structured representation, such as an abstract syntax tree (AST), and then input the AST into a natural language model. This can allow the model to better understand the structure and meaning of the code, and can be useful for tasks such as code completion or code formatting.

Bytecode, which is a low-level representation of code that is typically executed by a virtual machine, can also be processed using natural language models. One approach is to convert the bytecode into a higher-level representation, such as assembly code, and then input it into a natural language model. Another approach is to treat the bytecode as a sequence of tokens and input it into a sequence-to-sequence natural language model, which can be useful for tasks such as bytecode translation or bytecode summarization.

\subsubsection{Embeddings in \NLP}
An embedding is a dense, continuous-valued vector representation of a word or token. It is used to represent the meaning of the word or token in a numerical form that can be input into a machine learning model. Embeddings are commonly used in \NLP to represent words or tokens in a way that captures the semantic meaning and relationships between the words.

There are several reasons why embeddings are useful in \NLP. One reason is that they allow the model to handle large vocabularies more efficiently, as the model does not have to learn separate weights for each word in the vocabulary. Another reason is that embeddings can capture the relationships between words, such as synonymy and analogy, which can be useful for tasks such as language generation or translation. Finally, embeddings can also improve the generalization performance of the model by allowing it to handle out-of-vocabulary words or words that were not seen during training.

\section{\name Overview}\label{sec:models}
We begin by outlining the system and threat models before providing an overview of the key components of our proposed solution, which we refer to as \name.

\subsection{System Model}
Our system model considers a blockchain ledger that employs smart contracts and cryptocurrency assets, enabling traders and attackers to conduct transactions across various DeFi platforms such as exchanges, lending, leveraging. In this study, we specifically focus on \EVM-based blockchains.

\begin{itemize}
\item \textbf{Transaction and State Transition:} A blockchain ledger functions as a state machine replication, with its state denoted by $S$. Users define financial operations within a blockchain transaction, represented as $tx$, to request state transitions on the blockchain. The transaction serves as a state transition function that alters the ledger's state from $S$ to $S'$. In other words, $S' = tx(S)$.
\item \textbf{Smart Contract:} A smart contract is a piece of code that translates into one or more state transition functions, which can be activated by a transaction. Smart contracts can also trigger functions of other contracts.
\item \textbf{Blockchain Nodes:} A blockchain node may be assigned to one or more tasks: (i) transaction sequencing, determining the order of transactions within a block; (ii) block generation; (iii) data verification; and (iv) data propagation. The two prevalent types of blockchain nodes are:
\begin{itemize}
\item \textbf{Sequencer nodes}, also referred to as miners in PoW blockchains, validators in PoS blockchains, and block builders in PBS, encompass all four responsibilities mentioned above. Sequencers can insert, omit, and reorder transactions in the blocks they create within the boundaries set by the protocol.
\item \textbf{Ordinary nodes} solely handle blockchain data propagation and might also perform data verification.
\end{itemize}
\item \textbf{Transaction Propagation:} There are primarily two methods for propagating transactions from the transaction generator to the sequencer nodes:
\begin{itemize}
\item \textbf{Public Propagation:} Blockchain network protocols generally guide nodes on discovering and connecting to other nodes within the peer-to-peer (P2P) network. Transactions can be disseminated in the P2P network from the transaction generator, to the corresponding sequencer nodes, through ordinary nodes.
\item \textbf{Private Propagation:} \FaaS services enable DeFi traders to submit transactions directly to sequencer nodes, bypassing a broadcast on the P2P network. \FaaS may not be available on certain blockchains (e.g., Binance Smart Chain and Avalanche) and may be the only option on chains with a sole sequencer (e.g., Optimism). Ethereum presently supports both P2P and \FaaS propagation.
\end{itemize}
\item \textbf{Transaction Execution Trace:} A transaction execution trace documents the sequence of actions and state changes resulting from processing a transaction. Transaction traces can be represented in various ways, such as showcasing low-level OP codes or high-level DeFi operations.
\end{itemize}

\subsection{Threat Model}
We consider a computationally bounded adversary (denoted by $\mathcal{A}$) which is capable of executing transactions (i.e., performing actions) across a set of DeFi platforms. $\mathcal{A}$ may exploit vulnerabilities, in an attempt to alter a DeFi protocol's designed, expected state transition. Our threat model captures two different types of adversaries based on their capabilities:

\begin{itemize}
\item \textbf{Observable Adversaries}: These adversaries do not have the capability to hide their pending transactions from~\name. Their transactions are observable to our system either by \textit{(i)} broadcasting on P2P such that anyone can observe; or if \textit{(ii)}~\name is used by sequencer nodes, so even if the adversary uses private propagation (\FaaS), the sequencer will still be able to observe adversarial transactions.

\item \textbf{Hidden Adversaries}: These adversaries are capable of hiding their transactions from~\name until the adversarial transactions are finalized. Hidden adversaries achieve this by using a \FaaS system or similar method.
\end{itemize}

\subsection{\name Overview}

\name consists of three main components: a transaction tracer, a training module, and a detection module. The intuition behind this design choice is that normal transactions and attack transactions have different execution paths, and a powerful model can identify attack transactions based on their unusual execution paths.

\begin{enumerate}
\item The transaction tracer captures the execution trace of a transaction initiated by a user interacting with a \DApps. The trace includes the sequence of smart contract function calls, the associated input and output data, and provides a detailed view of the execution path. This trace is used as input for the detection module.
\item The training module uses a dataset of historical transactions to train a model using unsupervised or self-supervised methods. The goal of the training phase is to learn a model that can differentiate between normal and abnormal transaction execution traces. This allows \name to identify anomalies in real-time transactions.
\item The detection module applies the trained model to new transaction traces to generate a score based on the log-likelihood of the trace. A ranking or threshold-based method is then used to raise alarms for transactions with abnormal scores. This allows \name to detect potential attacks on the blockchain in real-time and prevent them from causing harm to the DeFi platform.
\end{enumerate}

\subsection{Motivating Examples}

We present two real-world examples that illustrate the potential benefits of using \name in the DeFi ecosystem.

\begin{itemize}
    \item \textbf{Motivating Example 1 (Observable Adversary):} Consider the attack on the Beanstalk project in April 2022. In a single transaction, the attacker borrowed one billion USD in cryptocurrency assets through Aave's flash loan, exchanged the borrowed assets for a $67\%$ stake in the Beanstalk project, and subsequently passed their proposal to withdraw the entire treasury. Etherscan received the adversarial transaction (\abbrEtherscanTx{0xcd314668aaa9bbfebaf1a0bd2b6553d01dd58899c508d4729fa7311dc5d33ad7}) on the public \PtP network approximately~$30$ seconds before its block confirmation, indicating that the attacker was an observable adversary.

    Table~\ref{tab:top1} shows that our evaluation ranks the adversarial transaction as the most abnormal among all historical interactions with the Beanstalk victim contract. This suggests that, if Beanstalk had utilized \name alongside a well-connected blockchain node within the P2P network, it would have had a 30-second window to detect and respond to the attack. In reaction to such an attack, Beanstalk could preemptively counter the adversary by initiating an emergency withdrawal of user funds or enforcing an emergency pause. This example demonstrates how \name can enable DeFi protocols to proactively detect and prevent malicious activities, ensuring the security and integrity of users' assets.

    \item \textbf{Motivating Example 2 (Hidden Adversary):} Consider the attack on the Revest Protocol in March 2022, during which approximately two million USD worth of tokens were stolen in four transactions. The root cause of the attack was a reentrancy vulnerability in a minting contract. The attack comprised four transactions confirmed between block \block{14465357} and block \block{14465427}, spanning roughly 70 blocks over a period of about 17 minutes. All transactions were propagated through a \FaaS relayer, specifically Flashbots.

    Suppose Revest Finance had deployed \name as an ordinary blockchain node without colluding with sequencers. Since the adversarial transactions were not visible on the \PtP network until they were confirmed, the attacker would be considered a hidden adversary.

    In this scenario, \name could only act as a fast retrospective attack detection tool. If \name detected the first adversarial transaction upon receiving block \block{14465357}, it could have potentially prevented the other three attacks. Our evaluation found that \name ranked the first adversarial transaction in Revest Finance as the most abnormal transaction. This example underscores the importance of having a tool like~\name, even if it is unable to observe pending transactions.
\end{itemize}

\section{\name Details}\label{sec:details}
We now proceed to elaborate on the details of \name. 

\subsection{\name Components}
\begin{figure*}[htb]
    \centering
    \includegraphics[width=\textwidth]{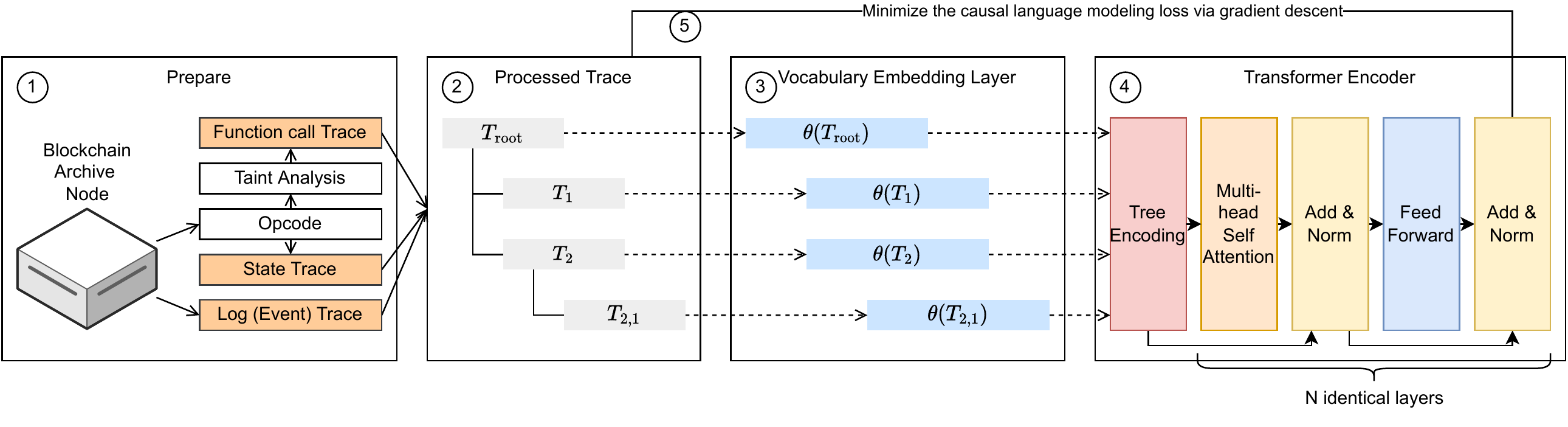}
    \caption{\name System Architecture. In step \circled{1} we extract the transaction trace from a blockchain archive node and augment the trace with additional data. We then process the trace in a graph structure in step \circled{2} and apply a vocabulary embedding layer in step \circled{3}. In step \circled{4} we apply a custom positional tree encoding, coupled with a loop to minimize the model loss in step \circled{5}.}
    \label{fig:system_architecture}
\end{figure*}

\name generates transaction representations for anomaly detection of blockchain transactions in real-time (cf.\ Figure~\ref{fig:system_architecture}). As such, \name takes as input a blockchain transaction (e.g., from the transaction pool), transforms it into a vector, extracts a vocabulary and applies a transformer in an effort to learn a probability distribution on an entire transaction's blockchain trace. \name's design consists of the following six components (cf. Section~\ref{sec:details} for details).
\begin{itemize}
    \item \textit{\circled{1} \ITR Construction}: Given a transaction, we construct a tree structured trace. This trace captures not only the smart contract function call dependencies, but also finer granularity information such as state accesses and event logs. Each node in the \ITR tree can be thought of as a step in a transaction execution process. An \ITR node may consist of a stream of bytes, which are further split into \ITR tokens in step \circled{2} (cf.~\ref{lst:itr} for an example of an \ITR construction).
    
    \item \textit{\circled{2} \ITR Tokenization}: Given domain specific heuristics, we first split all \ITR nodes into tokens, which then inform a custom vocabulary, drawn by the frequency distribution of those \ITR tokens. An \ITR token, for instance, can be a function name, event name, parameter type, parameter value, etc. To satisfy the requirements of step \circled{3} and  \circled{4}, we pre-process the \ITR trace according to our custom grammar.
    \item \textit{\circled{3} Vocabulary Embedding}: 
    Vocabulary embedding is a technique used to convert words or tokens in a text to numerical vector representations. It is commonly used in \NLP to input text data into machine learning models.

    To perform vocabulary embedding in the given process, a one-hot encoding function is first applied to the \ITR tokens, which converts the tokens into a binary vector representation where each position corresponds to a unique token and the value is either $0$ or $1$.

    The resulting vector representation of each token is the sum of three different embeddings: $E_{tok}$, $E_{tree}$, and $E_{ctx}$. $E_{tok}$ is the unique embedding of the token, which represents the token itself. $E_{tree}$ is an embedding that represents the position of the function call involving the token in the call graph, which forms a tree structure. $E_{ctx}$ is an embedding that represents the context in which the token appears, such as whether an address appears as a ``from'' or ``to'' address.

    By combining these three embeddings, the vocabulary embedding technique can capture both the unique characteristics of the tokens, their context within the trace, and their purpose. This can help the machine learning model to better understand the meaning and relationships between the tokens in the trace.

    \item \textit{\circled{4} Transformer Encoder}: 
    A processed transaction trace is initially converted using the prior vocabulary embedding layer, which associates the words or tokens in the trace with a dense vector representation. This representation is then fed into this step involving a transformer-based language model.

    As transformer-based models employ attention mechanisms to discern the relationships between input tokens, they necessitate a positional encoding to comprehend the relative location of each token in the input. It is important to consider that the input here is a transaction trace, which takes the form of a tree structure rather than a linear sequence of characters. Consequently, the graph structure is supplemented with positional tokens to preserve the tree information while providing the encoding to the transformer model. This enables the model to utilize its multi-head attention mechanism to learn the intricate relationships among the input tokens.

    \item \textit{\circled{5} Loss Minimization}:
    We apply the following steps to minimize the causal language modeling loss. We first define a causal language model loss function, which measures the difference between the predicted probability distribution of the next trace token in the sequence and the actual probability distribution of the next trace token. We then compute the gradient of the loss function with respect to the model parameters. We can then update the model parameters using the gradients and a learning rate with gradient descent. We repeat the above steps until the loss has reached a satisfactory minimum or a maximum number of iterations has been reached.

    \item \textit{\circled{6} Ranking-based intrusion detection}
    We adopt a ranking-based approach to intrusion detection.
    Given a \DeFi application, our \IDS ranks all transactions involving the application by the log-likelihood of their traces as computed by \name, and raises an alarm for the transactions with $\alpha\%$ lowest log-likelihood, i.e. the most abnormal transactions.
    The cost of running the \IDS can be adjusted by controlling the parameter $\alpha$.
    Out of~$124$ attacks in the dataset, \name identified~$20$ transactions as the most abnormal, 20 transactions as the second most abnormal, and 7 transactions as the third most abnormal.
\end{itemize}

\subsection{\ITR Construction}\label{subsec:itr_cons}
It is difficult to develop a scalable, just-in-time security system using the typical approach of performing execution analysis directly with low-level opcode traces because of the large space and time cost involved~\cite{noeth2009scalatrace}. Call traces and other high-level representations do not capture sufficient runtime information to allow for the generation of precise trace embeddings~\cite{wang2019learning}. By creating a new trace tree, which we refer to as an \ITR, we overcome the fundamental shortcomings of past techniques. \ITR is a high-level function that combines the three traces that follow into a single tree structure.

\begin{enumerate}
    \item During the execution of a transaction $tx$, the \textbf{call trace} captures the function call dependencies that occur. When a smart contract function calls into another function, the call trace records such inter- and intra-contract calls within its trace.
    
    We use a directed tree $Tree_{\text{call}}(tx) = (T_{\text{call}}, E_{\text{call}})$ to represent a call trace. Each node $t \in T_{\text{call}}$ in $Tree_{\text{call}}(tx)$ corresponds to an executed function call, including its corresponding runtime encoded call data and return data in bytes. We use the notation $t_{\text{call}}^1 \xrightarrow{e} t_{\text{call}}^2, e \in E_{\text{call}}$ to denote the edges in $Tree_{\text{call}}(tx)$, where a function $t_{\text{call}}^2$ (i.e., callee) is triggered by another function $t_{\text{call}}^1$ (i.e., caller). 
    
    \item Smart contracts can access and modify the volatile memory and persistent storage of a blockchain. A \textbf{state trace} records any read and write operations to the persistent storage that may occur during the execution of a transaction $tx$. In particular, a state trace consists of two sequences, each of which captures state accesses and state changes during the execution of $tx$.
    
    We denote a state trace of a transaction $tx$ with $Seqs_{state}(tx) = (T_{\text{state}}, E_{\text{state}})$.
    The first sequence $T_{\text{state}} = [t_{\text{state}}^1, \ldots, t_{\text{state}}^n]$ consists of state reads and writes details.
    If $t_{\text{state}}^i$ is a read operation, then it consists of a value tuple $[\text{key}, \text{val}]$, meaning that a function reads value $val$ from global storage (i.e., account storage) at position $key$ during $tx$'s execution. 
    Similarly, if $t_{\text{state}}^i$ is a write operation, then it consists of a two value tuple $[\text{key}, \text{val}]$, meaning that a function overrides the value at global storage position $key$ with $\text{val}$ during the execution of the function call.
    The second sequence $E_{\text{state}} = [t_{\text{call}}^1 \xrightarrow{e} t_{\text{state}}^1, \ldots, t_{\text{call}}^n \xrightarrow{e} t_{\text{state}}^n]$ captures which function call reads or writes the state.
    
    \item A \textbf{log trace} is a sequence of variables that the smart contract developer chooses to expose at runtime. 
    Logs help developers during contract debugging and extended data analytic tasks.
    
    Similar to the state trace, we use $Seqs_{log}(tx) = (T^{tx}_{\text{log}}, E^{tx}_{\text{log}})$ to denote a log trace.
    The first sequence $T_{\text{log}} = [t_{\text{log}}^1, \ldots, t_{\text{log}}^n]$ contains all emitted smart contract events. Each $t^{tx}_{\text{log}} = [\text{Contract}, \text{Event Hash}, \text{Data (in bytes)}] \in T_{\text{log}}$ consists of a contract address, an event's hash identifier, and the corresponding encoded data in bytes. 
    The second sequence $E_{\text{log}} = [t_{\text{call}}^1 \xrightarrow{e} t_{\text{log}}^1, \ldots, t_{\text{call}}^n \xrightarrow{e} t_{\text{log}}^n]$ captures which function call emits each log.
\end{enumerate}

\begin{lstlisting}[
float=t,
basicstyle=\ttfamily,
columns=fullflexible,
breaklines=true,
caption={An example of intermediate trace representation construction},
label={lst:itr}
]
CALL,from:0x99d...,to:0xe59...,data:c4f...
|- DELEGATECALL,from,0xe59...,to,0xe...,data,f39...
| |- READ, 0x95c..., 0x67a
| |- LOG1, 0x0b8..., 0x699
...
\end{lstlisting}

\textbf{Construction}: We construct the \ITR tree by traversing through state and log traces. We convert each state and log trace element into a child leaf node, and append the leaf node to $Tree_{\text{call}}(tx)$ (cf. Figure~\ref{fig:trace_construction}). 

\begin{figure}[tb]
    \centering
    \includegraphics[width=\columnwidth]{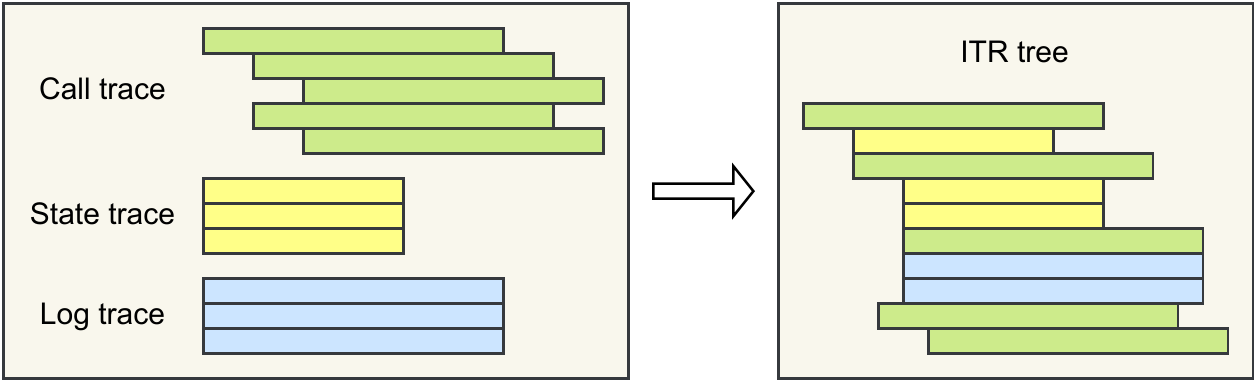}
    \caption{Abstract example of a Trace construction in \name.}
    \label{fig:trace_construction}
\end{figure}

\[
\begin{split}
    & Tree_{\text{ITR}}(tx) = (T_{ITR}, E_{ITR}) \\
    & = (T_{\text{call}} \bigcup T_{\text{state}} \bigcup T_{\text{log}}, E_{\text{call}} \bigcup E_{\text{state}} \bigcup E_{\text{log}})
\end{split}
\]

\subsection{\ITR Tokenization}\label{subsec:itr_tok}
In \NLP, tokenization is a common approach to identifying words that constitute a natural language sentence. Similarly, we apply tokenization to transform an \ITR trace tree to token sequences. We formally define an \ITR token as a string of arbitrary length. A token can, for example, represent a blockchain address, a function name, a log message, a storage key, a value, or a value type. We define our tokenization function $f_{\text{token}}(\cdot)$, taking as input either a call, state, or log \ITR node. In our grammar, the first two tokens of every node must be [START] followed by either one of the following three injected tokens: ([CALL], [STATE], [LOG]). The last token of every node must be [END]. In addition, we differentiate the start of input and output parameters with two added tokens ([INs], [OUTs]).

\begin{itemize}
    \item $f_{\text{token}}(T_{\text{call}}) =$ [START], [CALL], from, to, function hash, gas, value, [INs], input1 type, input1 value, $\dots$, [OUTs], output1 type, output1 value, $\dots$, [END]
    \item $f_{\text{token}}(T_{\text{state}}) =$ [START], [STATE], read / write, key, val, [END]
    \item $f_{\text{token}}(T_{\text{log}}) =$ [START], [LOG], contract address, event hash, value1 type, value1, $\dots$, [END]
\end{itemize}

We also choose to preprocess numerics to capture only the first two significant figures and the scale of numbers rather than the precise amounts (e.g., $1254 \xrightarrow{} 1300$). This is necessary to avoid vocabulary explosion because smart contracts frequently operate with big integers beyond $18$ decimals.

\paragraph{Break-Down of the Tokens}
We construct a vocabulary of $100$k tokens, as further elaborated in the evaluation, out of which 93,233 are Ethereum addresses and 6,759 are smart contract function signatures. Tokens which do not appear in our vocabulary are replaced with the [OOV] token.

\subsection{Local Token Embedding}
A local token embedding is the sum of three embeddings that respectively encode the identity, function call position, and local context of a token. For example, \name tokenizes the function call \texttt{addr1 -> addr2.func()} at the root of a call tree into three tokens: \texttt{[addr1]}, \texttt{[addr2]} and \texttt{[func]} (henceforth all tokens are enclosed by brackets). The local embeddings of the tokens \texttt{[addr1]} and \texttt{[addr2]}, both blockchain addresses, are $E_\texttt{[addr1]} + E_\text{root} + E_\text{src}$ and $E_\texttt{[addr2]} + E_\text{root} + E_\text{dst}$ respectively, where the embedding $E_\text{root}$ indicates that the position of the function call is at the root of the call tree, and the embedding $E_\text{src}$ and $E_\text{dst}$ are two local context embeddings added to the local embeddings of all source and destination addresses, without which the transformer encoder cannot possibly distinguish the role of the two addresses in the function call. The local embedding of the token \texttt{[func]}, a function signature, is $E_\texttt{[func]} + E_\text{root} + E_\text{func}$. The local context embedding $E_\text{func}$ is hardly necessary, as a function signature in general can occur only in one place in each function call. We can similarly generate local embeddings for function call parameters.
All embeddings so far are simply vectors retrieved from a lookup table.

In the example above, we use the embedding $E_\text{root}$ from a lookup table to indicate that a function call occurs at the root of a call tree. To faithfully encode the position of each function call, we need to associate a unique embedding to each function call position in a call tree. Without loss of generality, we binarize all call trees into binary trees, with each function call being a node in the tree. As a binary tree of depth $d$ has $O(2 d)$ unique positions, it is infeasible to construct a lookup table that stores the embedding of each function call position. However, each node in a binary tree can be uniquely identified by the path that leads to the node starting from the root of the tree, which is equivalent to a sequence of binary actions whether to visit the left ($L$) or right child ($R$). The embedding of a node can thus be the sum of embeddings of the actions taken at each step. Mathematically, the embedding of the action sequence $b_1, \ldots, b_n \in \{L, R\}^n$ is given by  $\sum_{i = 1}^n E_{i, b_i}$, where \[
  E_{i, b_i} = \begin{cases}
    E_{i, L} & b_i = L \\
    E_{i, R} & b_i = R
  \end{cases}
\]
Intuitively, $E_{i, b_i}$ indicates that the action $b_i$ is taken at step $i$. This approach only requires a lookup table consisting of the embeddings $E_{1, L}, E_{1, R}, \ldots, E_{D, L}, E_{D, R}$ where $D$ is the chosen maximum depth. Empirically, we did not observe the summation leading to any numerical instability during training, likely due to layer normalization in the transformer encoder.
Notably, the maximum depth of call trees that this embedding scheme can handle is not bounded by the embedding dimension, which differs from prior works \cite{shiv2019novel}.

\subsection{Contextual Token Embedding and Generative Pre-Training}\label{subsec:gpt}
Given a collection of local token embeddings, a transformer encoder can yield a collection of contextual embeddings for each token, which can carry context-dependent meanings that depend on the objective that they are tuned to optimize. For a collection of tokens that is sequentially ordered as $x_1, \ldots, x_n$, we maximize as its objective the log-likelihood of the collection that is factorized as $\log p(x_1, \ldots, x_n) = \sum_{i = 1}^n \log p(x_i|x_{< i})$, where the context $x_{< i}$ is $x_1, \ldots, x_{i - 1}$ for $i > 1$, and the context $x_{< 0}$ is a special symbol representing an empty sequence. Assuming that the local embeddings encode the position of each token, as is the case above, we can designate the contextual embedding of the token $x_{i - 1}$ to represent the context $x_{< i}$, and transform it into a categorical distribution over the vocabulary, which gives the probability of the next token $x_i$ occurring.

One way to transform an embedding into a categorical distribution over a vocabulary of size $n$ is to apply a linear transform followed by soft-max. Let $A$ be a nd matrix, the probability of a token \texttt{[tok]} occurring given the context $x_{< i}$ is given by \begin{align*}
  s & = A z_{i - 1} \in R^n	\\
  p(\texttt{[tok]}|x_{< i}) & = \exp(s^{(\#\texttt{[tok]})}) / \sum_{j = 1}^n \exp(s^{(j)})
\end{align*}
where \#\texttt{[tok]} is an integer between $1$ and $n$ that is uniquely assigned to the token \texttt{[tok]}. The notation $s^{(j)}$ denotes the $j$th component of the vector $s$, and $s^{(\#\texttt{[tok]})}$ thus denotes the $\#\texttt{[tok]}$th component of $s$.

The factorization, however, mandates the transformer encoder to derive the contextual embedding of each token from a different context, namely, the token itself and all the tokens that precede the token in the sequence. Fortunately, with proper attention-masking, we can proceed as if computing the contextual embedding of each token with the entire sequence as context, which is crucial when applying transformers to large-scale datasets.

Since a transformer works with any collection of tokens with a sequential ordering, we can apply it to nodes in call trees by linearizing call trees with breadth-first traversal. We leave other linearization schemes for future work. It is worth emphasizing that linearization only generates sequential orderings for log-likelihood factorization; the local token embeddings represent the position of function calls in the actual, not the linearized, call trees, as described above.

\subsection{Transformer Encoder}
We briefly describe in this section the architecture of transformer encoder and the technique of attention masking, and refer interested readers to the original paper~\cite{vaswani2017attention}.

On a high level, given a collection of vectors, a transformer encoder outputs a collection of vectors of equal size. This is done by applying $n$ modules of identical architecture and independent parameters. Each module is the composite of two submodules: a multi-headed self-attention module and a position-wise feedforward module. Given a collection of vectors $z_1, \ldots, z_n$, a self-attention module generates a query vector $q_i$, key vector $k_i$, and value vector $v_i$ for each vector $z_i$ with linear transforms: \[
    q_i = Q z_i \quad k_i = K z_i \quad v_i = V z_i
\]
The self-attention module then generates a vector $\hat{z}_i$ for each vector $z_i$ by aggregating all value vectors $v_1, \ldots, v_n$, weighted by the attention weights they receive from $z_i$: \[
    \hat{z}_i = \sum_{j = 1}^n \alpha_{i, j} v_j
\]
The attention weights are the inner products between key and query vectors, normalized by soft-max:
\[
    \alpha_{i, j} = \exp(q_i^T k_j) / \sum_{k = 1}^n \exp(q_i^T k_k)
\]
The soft-max normalization ensures that the attention weights for any $z_i$ are summed to $1$, i.e.\ forms a proper probability distribution. A multi-head self-attention module repeats the procedure multiple times with different parameters ($Q$, $K$, and $V$), and concatenates the resulting vectors into one.

The position-wise feedforward module applies a two layer neural network to each $\hat{z}_i$. The neural network consists of two linear layers interleaved with a nonlinear layer: \[
    \hat{z}_i \leftarrow A \phi(C \hat{z}_i + d) + b
\] where $A$ and $C$ are matrices, $b$ and $d$ are vectors, and $\phi$ is a nonlinear function.
The resultant embedding is added to the original embedding: \[
    z_i \leftarrow norm(z_i + FF(\hat{z}_i))
\] where the normalization operation ensures numerical stability during training, and $FF$ denotes the position-wise feedforward network. The aforementioned operations are iteratively applied 8 times with different parameters for each layer.

\subsection{Learning with stochastic gradient descent}
Given a function $f(\theta)$, we can minimize it by gradient descent by updating $\theta$ iteratively as follows: \[
    \theta \leftarrow \theta - \alpha \nabla_\theta f(\theta)
\] where $\alpha$ is a learning rate set as a hyperparameter.
In the case of generative pre-training, the function $f(\theta)$ is the joint likelihood \[
    \sum_{x \in X} \log p_\theta (x) = \sum_{x \in X} \sum_{i = 1}^{n_x} \log p_\theta (x_i | x_{< i})
\] where $X$ is the collection of transactions and $\theta$ is the parameters of the transformer encoder.
In practice, however, it is impractical to compute $f(\theta)$ exactly since the collection $X$, in our case, consists of nearly 68M transactions.
We therefore approximate $f(\theta)$ with \[
    \sum_{x \in \hat{X}} \log p_\theta (x) = \sum_{x \in \hat{X}} \sum_{i = 1}^{n_x} \log p_\theta (x_i | x_{< i})
\] where $\hat{X}$ is a \emph{mini-batch} of transactions randomly sampled from $X$.
Notably, if the samples are drawn independently at random from an identical distribution, the resulting approximation is an unbiased estimate of $f(\theta)$, thereby guaranteeing the convergence of gradient descent. This technique is referred to as stochastic gradient descent. The gradient of this approximation to $f(\theta)$ can be computed by deep learning frameworks such as PyTorch~\cite{paszke2017automatic}.

We can also accelerate the convergence of the iterative procedure by using second-order information of $f(\theta)$ in addition to its gradient by, for example, \[
    \theta \leftarrow \theta - \alpha H_f (\theta)^{-1} \nabla_\theta f(\theta)
\] where $H_f (\theta)$ is the Hessian matrix of $f(\theta)$. Although $H_f (\theta)$ is impractical to evaluate in practice due to the large number of parameters in transformer encoders, multiple approximation techniques exist that utilize some of the information without prohibitive computation costs. In this paper, we use the AdamW optimizer~\cite{loshchilov2017decoupled} with its default learning rate and momentum.

\section{DeFi Intrusion/Anomaly Detection System}
In this work, we focus on the task of automated and real-time detection of abnormal transactions, such as \DeFi attacks. Previous works~\cite{wu2021defiranger,qin2022quantifying} rely on costly and time-consuming manual feature extraction and modeling, and then propose custom heuristics to identify specific attacks. As a result, without significant effort, existing approaches cannot be scaled and generalized across different types of \DApps protocol designs on various blockchains. To overcome this limitation, we propose to automate the \IDS process by performing transaction level anomaly detection. Intuitively, adversarial transactions should be semantically distinguishable from benign transactions within and across \DApps.

\subsection{Data}
We extend a dataset comprised of all transactions that involve 124 previously compromised \DeFi applications on Ethereum~\cite{zhou2022sok}. Our dataset shows an average increase in incident frequency from 3.1/month in 2020 to 8.5/month in 2022 ($2.74\times$). The most common incident causes are smart contract Layer ($42\%$, e.g., reentrancy attack), protocol layer ($40\%$, e.g., price oracle manipulation), and auxiliary layer ($30\%$, e.g., honeypot) vulnerabilities. Our dataset consists of a total of~\TotalNumTxs transactions, spanning a period of~\DataNumOfDays days, starting from block~\DataStartBlock (\DataStartDate) and ending at block~\DataEndBlock (\DataEndDate), and was constructed as outlined in Section~\ref{subsec:itr_cons} and Section~\ref{subsec:itr_tok}. We chose to use this dataset because it includes known instances of compromised DeFi applications, providing a ground truth for our analysis. Our approach is unsupervised/self-supervised learning, which allows us to evaluate the effectiveness of~\name just using this dataset with previously exploited platforms.
We leave it as future work to enhance the dataset by including transactions from non-compromised applications, and focus on previously compromised applications in this paper. 

We pre-trained the transformer encoder as in Section~\ref{subsec:gpt} to maximize the joint likelihood of the traces of the~\TotalNumTxs transactions. With the maximum likelihood as its sole objective, generative pre-training needs no other data apart from the traces of the transactions in the dataset. For example, we do not need labels of whether a transaction is benign or anomalous, as in a supervised learning setting. 

When evaluating the intrusion detection capability of the pre-trained transformer encoder (the \IDS), we used all transactions in our dataset as the ground truth, assuming that only transactions tagged as malicious were malicious, while the rest of the transactions in the dataset are considered benign. It is important to note that our dataset has a limitation in that it is possible that other undiscovered attacks may have occurred prior or after the known attacks, which may affect the overall evaluation of the \IDS performance. Additionally, it is worth noting that each attack may involve more than one malicious transaction, some of which may appear benign but in fact prepare for the attack. Therefore, our approach may not be able to identify any anomalies based on the transaction traces of these transactions, which is another limitation of our study.

We evaluate the \IDS for each smart contract independently, as described below.

\subsection{Detection Methodology}
We adopt a ranking-based approach to intrusion detection. Recall that given a linearized transaction trace $x_1, \ldots, x_n$, a transformer encoder coupled with a linear transform and soft-max can yield a sequence of conditional log-likelihoods $\log p(x_1|x_{< 0}), \ldots, p(x_n|x_{< n})$, the sum of which is the log-likelihood of the trace. Given a DeFi application, \name ranks all transactions involved with the application by the log-likelihood of their traces, and raises an alarm for the $\alpha\%$ most abnormal transactions, where $\alpha$ is an adjustable parameter. Intuitively, increasing $\alpha$ makes it more likely for the \IDS to detect attacks, but also raises the likelihood of false positives.

\subsection{Implementation Details} 
We pack traces into min-batches to speed up training. Each mini-batch consists of 32 traces, each with 512 tokens. The mini-batch size is chosen to maximize GPU utilization (we use an A100 SXM4 GPUs with 40 GB memory). As most traces in the dataset yield far less than 512 tokens after tokenization, we concatenate as many short traces as possible into one trace that is no longer than 512, and apply proper attention masking to ensure that the log-likelihood of each trace is computed as if it is the only trace. Each mini-batch therefore contains a variable number of traces.

The dataset contains 5M unique Ethereum addresses and 5M unique Ethereum function signatures. It is infeasible to store all their embeddings on a GPU due to GPU memory constraints. We therefore store the embeddings of the most frequent 100K tokens in the GPU, and the embeddings of the rest in RAM. Embeddings in the GPU are updated synchronously as the rest of the parameters in the transformer encoder. Embeddings in RAM are updated by a parameter server \cite{li2014communication} using asynchronous stochastic gradient descent. This architecture enables us to store significantly more embeddings. This approach is highly efficient as the embeddings of rare tokens are rarely accessed. Without further optimization, the proposed \IDS attains a batched throughput of $2284 \pm 289$ transactions per second on a single NVIDIA A100 GPU. We use the AdamW optimizer \cite{loshchilov2017decoupled} with its default learning rate and momentum.

\section{Evaluation}\label{sec:evaluation}
This section presents an evaluation of \name as an \IDS for detecting \DeFi attacks. We begin by analyzing \name's performance under various alarm threshold configurations. We then employ common evaluation metrics such as Precision, Recall, and F-score to further assess the system's effectiveness. Additionally, we compare \name against several benchmark approaches to demonstrate its capability in detecting a diverse range of attacks. For readers interested in a more advanced, \IDS-specific metric for comparison with related works, the Intrusion Detection Capability Score is evaluated in the appendix (cf.\ Appendix~\ref{app:cid}).

\subsection{Assumption}
In the following we assume that for incidents involving multiple adversarial transactions, it is sufficient for \name to detect just one transaction in the sequence, rather than all of them. This is based on the premise that, upon detecting an abnormal transaction, DeFi protocol operators can take immediate action to prevent further harm, such as activating an emergency pause on the entire protocol. Research indicates that approximately $50\%$ of attacked DeFi protocols already have such a pause mechanism in place~\cite{zhou2022sok}.

\begin{figure*}[tb]
    \centering
    \resizebox{\textwidth}{!}{%
        \begin{tabular}{l|ccccc|ccc}
        \toprule
        \multirow{2}{*}{\makecell{Dataset Size (the total number of transactions \\ interacting with the vulnerable smart contract)}} 
            & \multicolumn{5}{c|}{Percentage Ranking Alarm Threshold (\%)} & \multicolumn{3}{c}{Absolute Ranking Alarm Threshold} \\
            & $\leq0.01\%$ & $\leq0.1\%$ & $\leq0.5\%$ & $\leq1\%$ & $\leq10\%$ & top-1 & top-2 & top-3 \\
        \midrule
        \midrule
        0 - 99 txs (32 attacks, 28\% of dataset)      
            & -         & -         & -       & -         & 5 (16\%) &   7 (22\%) & 20 (63\%) & 23 (72\%) \\
        Average false positive rate 
            & -         & -         & -       & -         & 8.18\%   &   0\%      & 14.8\%    & 28.3\%    \\
        Average number of false positives   
            & -         & -         & -       & -         & 5.1      &   0        & 1         & 2         \\
        \midrule
        100 - 999 txs (38 attacks, 33\% of dataset)   
            & -         & -         & 8 (21\%) & 12 (32\%) & 28 (74\%) &  7 (18\%) & 12 (32\%) & 15 (39\%) \\
        Average false positive rate 
            & -         & -         & 0.24\%   & 0.71\%    & 9.65\%    &  0\%      & 0.46\%    & 0.81\%    \\
        Average number of false positives  
            & -         & -         & 1.5      & 3.5       & 39.4      &  0        & 1         & 2         \\
        \midrule
        1000 - 9999 txs (17 attacks, 15\% of dataset) 
            & -         & 6 (35\%)  & 9 (53\%)   & 11 (65\%)  & 13 (76\%) & 4 (24\%) & 7 (41\%) & 7 (41\%) \\
        Average false positive rate 
            & -         & 0.054\%   & 0.45\%     & 0.95\%     & 9.96\%    & 0\%      & 0.049\%  & 0.098\%  \\
        Average number of false positives  
            & -         & 1.4       & 11.5       & 23.7       & 324.5     & 0        & 1        & 2        \\
        \midrule
        10000 + txs (29 attacks, 25\% of dataset)      
            & 2 (7\%)   & 7 (24\%)  & 16 (55\%)  & 18 (62\%)  & 21 (72\%) & 2 (7\%)  & 3 (10\%) & 4 (14\%) \\
        Average false positive rate 
            & 0.007\%   & 0.097\%   & 0.50\%    & 1\%        & 10\%      & 0\%      & 0.004\%  & 0.008\%  \\
        Average number of false positives  
            & 2.5       & 120.1     & 429.9     & 819.6      & 7302.1    & 0        & 1        & 2        \\
        \midrule
        \midrule
        Overall                                                 
            & 2 (2\%)   & 13 (11\%) & 33 (28\%) & 41 (35\%)  & 67 (58\%) & 20 (17\%) & 42 (36\%) & 49 (42\%) \\
        Average false positive rate 
            & 0.007\%   & 0.077\%   & 0.42\%    & 0.90\%     & 9.71\%    & 0\%       & 7.19\%    & 13.5\%    \\
        Average number of false positives  
            & 2.5       & 65.3      & 211.9     & 367.2      & 2368.5    & 0         & 1         & 2        \\
        \bottomrule
        \end{tabular}
    }
\caption{
This table presents the performance of \name under various alarm threshold configurations, organized by the number of transactions interacting with the vulnerable smart contracts. For example, with an alarm threshold of $\leq 0.01\%$, our method detects $24\%$ of the attacks within the $10000+$ transaction range, with an average false positive rate of $0.097\%$. The results indicate that using a lower alarm threshold enables the detection of a higher percentage of attacks, albeit at the cost of an increased false positive rate. Notably, the efficacy of the alarm threshold varies across different dataset sizes, emphasizing the need to select a suitable threshold based on the specific attributes of the smart contract under investigation.
}
\label{fig:threshold}
\end{figure*}

\subsection{Alarm Threshold and Metrics}
We assess the performance of \name by analyzing various alarm threshold configurations. The alarm threshold is a system parameter that users must select for \name. It determines the sensitivity of an \IDS and is defined as the likelihood of a transaction being deemed abnormal. Transactions that fall below this threshold will generate an alarm. As an example, when the alarm threshold is set to $1\%$, the \IDS will raise alarms for the $1\%$ least likely, or most abnormal, transactions interacting with a smart contract.

We consider two different types of alarm thresholds in this paper: \textit{(i)} Percentage ranking alarm threshold, which is expressed as a proportion of transactions in the dataset; and \textit{(ii)} Absolute ranking alarm threshold, which corresponds to a fixed number of top-ranked transactions.

Note that the detection thresholds granularity depends on the dataset sizes. For example, if a smart contract has~$100$ interacting transactions, the alarm threshold would be~$1\%$ even when inspecting a single transaction, which is the best possible scenario. This limitation arises due to the inherent constraints of the dataset size and should be considered when interpreting the results in Figure~\ref{fig:threshold}.

\subsection{Effectiveness}

\begin{figure}[tb]
  \centering
  \resizebox{\columnwidth}{!}{%
  \begin{tabular}{lllr}
    \toprule
         Victim Name &                            \makecell{Victim \\ Contract} & \makecell{Application\\Categories} & \makecell{Damage\\(in USD)} \\
    \midrule
           Beanstalk & \abbrEtherscanAddress{0xc1e088fc1323b20bcbee9bd1b9fc9546db5624c5} &             Stablecoin &  181,500,000 \\
               MonoX & \abbrEtherscanAddress{0x66e7d7839333f502df355f5bd87aea24bac2ee63} &                    DEX &   31,133,333 \\
     PopsicleFinance & \abbrEtherscanAddress{0xd63b340f6e9cccf0c997c83c8d036fa53b113546} &          Yield farming &   20,700,000 \\
    PrimitiveFinance & \abbrEtherscanAddress{0x9daec8d56cdcbde72abe65f4a5daf8cc0a5bf2f9} &            Derivatives &   13,000,000 \\
        PunkProtocol & \abbrEtherscanAddress{0x929cb86046e421abf7e1e02de7836742654d49d6} &                 Others &    8,950,000 \\
        VisorFinance & \abbrEtherscanAddress{0xc9f27a50f82571c1c8423a42970613b8dbda14ef} &                 Others &    8,200,000 \\
            DAOMaker & \abbrEtherscanAddress{0xd6c8dd834abeeefa7a663c1265ce840ca457b1ec} &                 Others &    4,000,000 \\
            DAOMaker & \abbrEtherscanAddress{0x933fb39d2b0f110e6e83f62c4fbcaebfd3142a13} &                 Others &    4,000,000 \\
                DODO & \abbrEtherscanAddress{0x051ebd717311350f1684f89335bed4abd083a2b6} &                    DEX &    3,800,000 \\
                DODO & \abbrEtherscanAddress{0x509ef8c68e7d246aab686b6d9929998282a941fb} &                    DEX &    3,800,000 \\
          CheeseBank & \abbrEtherscanAddress{0x833e440332caa07597a5116fbb6163f0e15f743d} &           Digital Bank &    3,300,000 \\
                dydx & \abbrEtherscanAddress{0x53773fe5ff4451c896127dd2c91b8de7ac51ba2c} &            Derivatives &    2,211,000 \\
       RevestFinance & \abbrEtherscanAddress{0xe952bda8c06481506e4731c4f54ced2d4ab81659} &                 Others &    2,005,000 \\
           BTFinance & \abbrEtherscanAddress{0x3ec4a6cfe803ee84009ce6e1ecf419c9cb1e8af0} &          Yield farming &    1,600,000 \\
        VisorFinance & \abbrEtherscanAddress{0x65bc5c6a2630a87c2b494f36148e338dd76c054f} &                 Others &      975,720 \\
          WildCredit & \abbrEtherscanAddress{0x7b3b69eab43c1aa677df04b4b65f0d169fcdc6ca} &                Lending &      650,000 \\
         SharedStake & \abbrEtherscanAddress{0xa23179be88887804f319c047e88fdd4dd4867ef5} &                 Others &      500,000 \\
               88mph & \abbrEtherscanAddress{0x2165b3800b17224de39303c240a41064179db0a6} &                Lending &      100,000 \\
           SanshuInu & \abbrEtherscanAddress{0x35c674c288577df3e9b5dafef945795b741c7810} &                 Others &      100,000 \\
     KlondikeFinance & \abbrEtherscanAddress{0xacbdb82f07b2653137d3a08a22637121422ae747} &       Synthetic assets &       22,116 \\
    \bottomrule
    \end{tabular}
    }
    \caption{The $20$ attacks ranked by \name \IDS as the most abnormal transaction that interacted with the respective victim contract. \name successfully identifies the most abnormal transaction for $18$ unique \DeFi protocols across various application categories, with the total damage value amounting to over $276$ million USD.}
    \label{tab:top1}
\end{figure}

To provide a comprehensive analysis, we evaluate the effectiveness of \name across various alarm threshold settings (i.e., $0.01\%$, $0.1\%$, $0.5\%$, $1\%$, $10\%$ for percentage ranking alarm threshold, and top-1, top-2, top-3 for absolute ranking alarm threshold) and different dataset sizes (i.e., 0-99, 100-999, 1000-9999, and 10000+ transactions). Figure~\ref{fig:threshold} presents the performance results for each alarm threshold level, including the following metrics: \textit{(i)} The number of detected attacks (and the corresponding percentage of total attacks); \textit{(ii)} The average false positive rate; and \textit{(iii)} The average number of false positives.

As demonstrated in Figure~\ref{fig:threshold}, \name is proficient at identifying abnormal transactions, ranking 49 out of the 124 attacks ($42\%$) among the top-3 most abnormal transactions interacting with their respective victim contracts. In particular, the top-1 most abnormal transactions and the associated damages incurred by their victims are displayed in Figure~\ref{tab:top1}. \name effectively detects the most abnormal transaction for $18$ unique \DeFi protocols across various application categories, with the total damage value exceeding $276$ million USD. In more detail, \name ranked 20 adversarial transactions as the most abnormal transactions involving their victim contracts, 22 as the second least likely, and 7 as the third least likely. The top-2 and top-3 most abnormal transactions can be found in Appendix~\ref{app:top_2_3}.

Note that our data suggests that there is a trade-off between the false positive rate and how many attacks \name captures. For example, consider the case of smart contracts with 1000-9999 transactions. With an alarm threshold of $0.1\%$, \name detects $35\%$ of attacks while maintaining a false positive rate of $0.054\%$. Increasing the threshold to $0.5\%$ improves the detection rate to $53\%$ with a corresponding false positive rate of $0.45\%$. Although the detection rate has increased by 18 percentage points, the false positive rate has also increased by approximately~$8.33$ times. This demonstrates that there is a trade-off between improved detection capabilities and higher false positive rates. 

Our findings indicate that the performance of \name is positively correlated with the size of the dataset. For instance, at the same alarm threshold of~$0.5\%$, our model demonstrates better performance on larger transaction sets. It detects~$55\%$ of the attacks within the~$10000+$ transaction range,~$53\%$ for the~$1000-10000$ range, and only~$21\%$ for the~$100-1000$ range. This observation suggests that \name may benefit from larger datasets, as it allows for more accurate identification of attacks while maintaining an acceptable false positive rate. 

\subsection{Practicality}
Our analysis highlights the practicality of \name when handling a large number of transactions. For instance, \name demonstrates its best performance on historical data for \DeFi applications with a transaction history ranging from $1,000$ to $9,999$. Using a $0.01\%$ alarm threshold, \name captures $35\%$ of the attacks while maintaining an average false positive rate (FPR) of $0.054\%$, resulting in only $1.4$ false positive transactions on average. Additionally, in \DeFi applications with over $10,000$ transactions and a $0.01\%$ alarm threshold, \name successfully detects $24\%$ of the attacks with an average FPR of $0.097\%$, corresponding to $120.1$ false positive transactions on average.

In the context of popular \DeFi applications processing $100$ transactions per day, a $0.1\%$ \FPR generates one alert approximately every 10 days. This showcases that \name can provide a manageable number of alerts for further investigation, making it particularly suitable for high-volume transaction environments.

\subsection{Performance}
\name exhibits real-time capabilities, achieving an average batch throughput of $2,284 \pm 289$ transactions per second and taking an average of $0.16 \pm 0.3$ seconds to rank a single transaction.

\subsection{Precision, Recall and $F_1$ Scores}
This section aims to visualize traditional machine learning metrics for our tool, \name, and provide insights into its performance. We begin by providing definitions for the metrics and key terms:

\begin{itemize} 
    \item True Positives (TP): Number of correctly identified adversarial transactions.
    \item False Positives (FP): Number of non-adversarial transactions incorrectly classified as adversarial.
    \item False Negatives (FN): Number of adversarial transactions incorrectly classified as non-adversarial.
    \item Precision ($\frac{\text{TP}}{\text{TP} + \text{FP}}$) is the fraction of adversarial transactions among alarms raised.
  \begin{align*}
      \text{Prec} = \frac{\text{number of adversarial txs \name captures}}{\text{number of alarms \name raises}}
  \end{align*}
  \item Recall ($\frac{\text{TP}}{\text{TP} + \text{FN}}$) is the fraction of attacks that are detected.
  \begin{align*}
        Recall = \frac{\text{number of adversarial txs \name captures}}{\text{total number of adversarial txs}}
  \end{align*}
  \item $F_1$-score is the harmonic mean of precision and recall.
  \begin{align*}
      F_1 = \frac{2 \cdot \text{Prec} \cdot \text{Recall}}{\text{Prec} + \text{Recall}}
  \end{align*}
\end{itemize}

\begin{figure}[tb]
    \centering
    \includegraphics[width=\columnwidth]{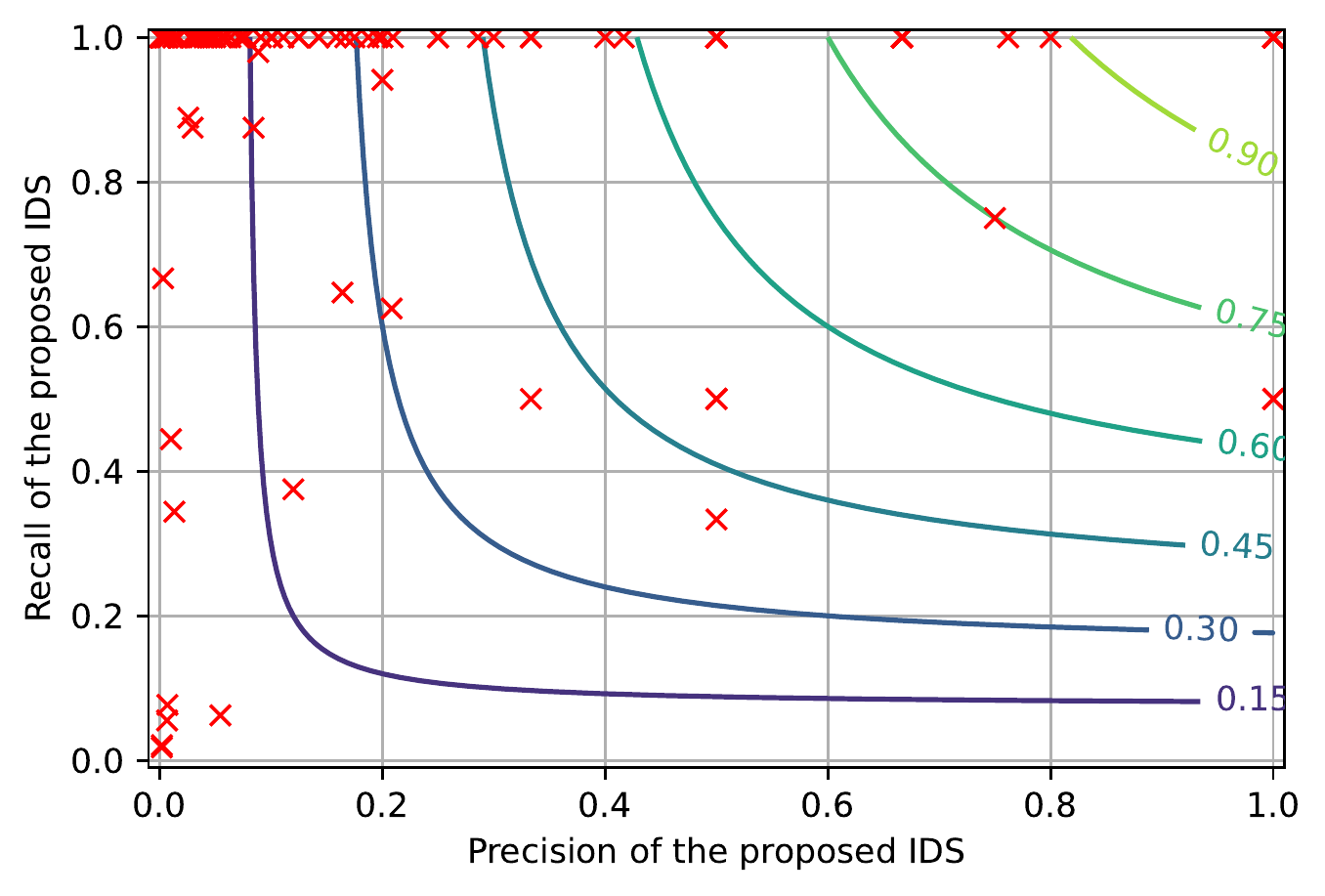}
    \caption{The figure presents a detailed evaluation of the performance of the proposed \name \IDS by analyzing its $F_1$ score for each previously compromised \DeFi application in our dataset. Each data point represents the precision and recall achieved by the \name \IDS for a specific \DeFi application, thereby highlighting its effectiveness in detecting various types of attacks. The contour lines depict the $F_1$-score as a bivariate function of the precision and recall of the IDS, offering a comprehensive view of the model's performance across different applications and showcasing its ability to provide real-time intrusion detection for blockchains such as Ethereum.}
    \label{fig:f1}
\end{figure}

Figure~\ref{fig:f1} presents the best F1-score  that \name can achieve, and the corresponding precision and recall, for each previously compromised \DeFi application in our dataset. We observe that~$112$~($96\%$) of our data exhibits a recall above $0.8$, while~$64$~($55\%$) has a precision of below $20\%$. The average F1 score is only $37.6\%$. However, it is important to note that while the traditional metrics may not appear as impressive, it does not imply that our system is not useful. The discrepancy can be attributed to the data imbalance issue and the so-called base rate fallacy problem.

Consider a scenario with a dataset containing one million transactions and only one adversarial transaction. An \IDS with an alarm triggering threshold of $0.01\%$ (raises in total $100$ alarms) can only attain a precision of $1\%$. In this case, the recall would be $100\%$ because the attack is detected. The F1-score, which equally weighs precision and recall, would be approximately $1.96\%$. Given that the cost of an attack is tremendous, in such cases, a high detection rate is desirable as it indicates that the \IDS is better at detecting intrusions, even if it comes at the cost of a higher false alarm rate.

An alternative approach for visualization is using the $F_\beta$-score to place greater emphasis on recall than precision (refer to the $F_{10}$ score in Appendix~\ref{app:f10}). However, this approach is subjective unless the value of $\beta$ is thoroughly justified quantitatively. For a more quantitative method to evaluate \name, we recommend referring to Appendix~\ref{app:cid}, where we discuss the Intrusion Detection Capability Score (CID score).

\subsection{Impact of Flash Loans on \name's Efficiency}
We proceed to examine the effectiveness of \name in detecting specific attack types, particularly focusing on whether the transaction utilizes a flash loan or not. Gaining a more profound understanding of this aspect is crucial for assessing the efficacy of \name in identifying attacks employing various strategies. We find that out of the 124 attacks in our dataset, approximately 34 (27\%) employed a flash loan. We can break down \name's performance on these attacks as follows:

\begin{enumerate}
    \item In attacks involving a flash loan, \name ranks the adversarial transaction among the top-3 most abnormal transactions for 16 out of 34 cases (47\%).
    \item In attacks without a flash loan, \name ranks the adversarial transaction among the top-3 most abnormal transactions for 31 out of 90 cases (34\%).
    \item The current results indicate that \name has a higher success rate in detecting flash loan attacks while also exhibiting its capability to detect non-flash loan attacks.    
\end{enumerate}

\name performs slightly better on flash loan attacks (47\% vs 34\%), suggesting it considers other factors in the trace, not just naively classifying flash loans as attacks.

\subsection{Benchmark 1~---~Comparison with a Doc2Vec Model}
To evaluate \name against a naive baseline, we implement a transaction ranker using a combination of doc2vec and Gaussian mixture models. The doc2vec model treats the flattened transaction trace as a document for analysis. This baseline ranks 8 attacks in our attack dataset as the most abnormal transactions interacting with the victim contracts, 11 as the second least likely, and 5 as the third least likely. In contrast, \name ranks 20 attacks as the most abnormal transactions, 22 as the second least likely, and 7 as the third least likely, showcasing its superior performance. For a detailed explanation of the technical aspects, we refer interested readers to Appendix~\ref{app:baseline}, where the Gaussian mixture model parameters are estimated using Expectation-Maximization, and the number of clusters is chosen by minimizing the Bayesian information criteria.

\subsection{Benchmark 2~---~Comparison with a Trace Length Heuristic}
Based on the observation that malicious transactions often have abnormally long traces, we develop a heuristic-based \IDS that ranks transactions interacting with a contract according to their trace lengths and flags those with the longest traces. This heuristic system identifies 20 attacks within the top-3 ranked transactions in our dataset, among which 18 are also ranked in the top-3 by \name. Overall, \name ranks 49 attacks in our attack dataset as top-3 abnormal transactions, demonstrating enhanced performance compared to the heuristic-based baseline. Importantly, \name maintains its ability to detect these attacks even if their traces are deliberately shortened to evade the baseline system, as it only examines the first 512 elements of each trace.

\subsection{Discussion}
The effectiveness of \name is influenced by the choice of the detection threshold. Selecting an appropriate detection threshold should consider not only the performance of the \IDS, but also the risk appetite and cost-benefit trade-off for the DeFi protocol operator. In practice, DeFi protocol operators ought to balance their risk tolerance and cost-benefit considerations when choosing a suitable detection threshold. They may also contemplate employing multiple thresholds for different growth stages, such as implementing a higher threshold for a contract's initial transactions and a lower threshold for transactions involving substantial asset amounts. We leave it to future work to automatically suggest alarm thresholds based on applications and growth stages. Furthermore, \name can be combined with orthogonal security measures, including smart contract auditing and whitelisting, establishing a comprehensive security framework for DeFi protocols.

\section{Related Works}
\begin{figure*}[t]
    \centering
    \resizebox{\textwidth}{!}{%
    \begin{tabular}{@{}ccccc@{}}
    \toprule
    \makecell{Technique} & \makecell{Assumed Prior Knowledge} & \makecell{Searchspace Unrestricted\\From Vulnerability Patterns} & \makecell{Real-Time\\Capable} & \makecell{Application\\Agnostic} \\\midrule
    
    \multicolumn{5}{c}{\clg{Rank based -- the goal is to find all unexpected execution patterns, implicitly capturing vulnerabilities}} \\ \midrule
    \name (this paper) 
        & All historical transactions  & Unrestricted & \fc (0.16s) & \fc \\ \midrule
    
    \multicolumn{5}{c}{\clg{Reward based -- the goal is to extract financial revenue, implicitly capturing vulnerabilities}} \\ \midrule
    APE~\cite{qin2023imitation} 
        & N/A & Only profitable patterns  & \fc (0.07s) & \fc \\
    Naive Imitation~\cite{qin2022quantifying}
        & N/A & Only profitable patterns  & \fc (0.01s) & \fc \\
    DeFiPoser~\cite{zhou2021just} 
        & DApp models & \makecell{Only profitable patterns \\ + Limited by the DApp models} & \fc (5.93s) & \ec \\ \midrule
    
    \multicolumn{5}{c}{\clg{Pattern based -- the goal is to match / classify predefined known vulnerability patterns with rules (including machine learning methods)}} \\ \midrule
    Pattern based dynamic analysis~\cite{rodler2018sereum,wu2021defiranger,zhang2020txspector}      
        & Rule                                      & Limited by the rule  & \fc & \hc \\
    Pattern based fuzzing~\cite{ferreira2021confuzzius,wang2020oracle,grieco2020echidna,wustholz2020harvey,he2019learning,nguyen2020sfuzz}               
        & Rule + ABI / DApp models                  & Limited by the rule  & \hc & \hc \\
    Pattern based symbolic execution~\cite{conkas,albert2018ethir,torres2019art,he2019learning,nikolic2018finding,mossberg2019manticore,wang2019detecting,torres2018osiris,luu2016making,bose2022sailfish,nguyen2021sguard,so2021smartest}   
        & Rule + Source code / Bytecode             & Limited by the rule  & N/A & \hc \\
    Pattern based static analysis~\cite{ghaleb2022etainter,brent2020ethainter,tsankov2018securify,contro2021ethersolve,schneidewind2020ethor,grech2018madmax,wang2019detecting,rodler2018sereum,feist2019slither,tikhomirov2018smartcheck}       
        & Rule + Source code / Bytecode             & Limited by the rule  & N/A & \hc \\ \midrule
    
    \multicolumn{5}{c}{\clg{Proof based -- the goal is to prove that a set of smart contracts meet specific security properties}} \\ \midrule
    Formal verification~\cite{azzopardi2018monitoring,frank2020ethbmc,he2019learning,grishchenko2018ethertrust}
        & \makecell{Formal security properties \\+ Source code / DApp models} & \makecell{Limited by the \\ security properties}  & N/A & \hc \\
    \bottomrule
    \end{tabular}%
    }
    \caption{Systematization of intrusion detection / prevention techniques. Unlike reward-based approaches, \name employs an unrestricted search space, enabling it to identify unexpected execution patterns instead of focusing solely on profitable vulnerabilities. In contrast to pattern-based techniques (dynamic analysis, fuzzing, symbolic execution, and static analysis),~\name does not rely on predefined rules or patterns, which allows it to detect a broader range of anomalies. Furthermore, \name is capable of real-time analysis, a feature not present in pattern-based symbolic execution or static analysis methods.}
    \label{tab:comparison}
\end{figure*}

\subsection{Smart Contract Intrusion Prevention and Detection}
Intrusion detection and prevention are essential components in the realm of decentralized finance security research. Table~\ref{tab:comparison} compares our proposed method with alternative approaches.

\begin{enumerate}
    \item Qin~\etal~\cite{qin2022quantifying,qin2023imitation} introduce two generalized imitation attack methods utilizing dynamic program analysis to automatically observe, copy, and synthesize profitable transactions from the \PtP network. While these methods find profitable transactions, they do not determine whether these profitable transactions are abnormal or not.
    \item DeFiPoser~\cite{zhou2021just} uses logical DeFi protocol models and a theorem prover (e.g., Z3) or the Bellman-Ford-Moore algorithm to create profitable DeFi transactions. It operates in real-time or offline. However, it also does not distinguish attacks and relies heavily on provided protocol models.
    \item Fuzzing involves providing generated inputs to a smart contract to uncover vulnerabilities. The idea is to test the contract's behavior in unexpected situations, which can reveal potential security issues. Fuzzing has been shown to be effective in detecting vulnerabilities in smart contracts, but it has limitations, such as the lack of coverage of all possible code paths~\cite{ferreira2021confuzzius,wang2020oracle,grieco2020echidna,wustholz2020harvey,he2019learning,nguyen2020sfuzz}.
    \item Symbolic execution involves evaluating a smart contract's code with symbolic inputs, rather than concrete values. The goal is to explore all possible code paths and uncover potential vulnerabilities. Symbolic execution has been used to detect various types of vulnerabilities, such as reentrancy attacks and integer overflow/underflows~\cite{conkas,albert2018ethir,torres2019art,he2019learning,nikolic2018finding,mossberg2019manticore,wang2019detecting,torres2018osiris,luu2016making,bose2022sailfish,nguyen2021sguard,so2021smartest}.
    \item Formal verification involves using mathematical methods to prove that a smart contract meets certain security properties~\cite{azzopardi2018monitoring,frank2020ethbmc,he2019learning,grishchenko2018ethertrust}. The idea is to formally prove that the contract's code is correct, which can provide a higher level of assurance than testing.
    \item Static analysis involves analyzing a smart contract's code without executing it. The goal is to uncover potential vulnerabilities by examining the contract's structure and control flow. Static analysis can be used to detect various types of vulnerabilities, such as uninitialized variables and unsafe function calls~\cite{ghaleb2022etainter,brent2020ethainter,tsankov2018securify,contro2021ethersolve,schneidewind2020ethor,grech2018madmax,wang2019detecting,rodler2018sereum,feist2019slither,tikhomirov2018smartcheck}.
    \item Dynamic analysis involves executing a smart contract and monitoring its behavior. The goal is to uncover potential vulnerabilities by observing the contract's behavior in a real-world environments~\cite{rodler2018sereum,wu2021defiranger,zhang2020txspector}.
\end{enumerate}

While the aforementioned techniques are effective in identifying vulnerabilities, they are not typically considered as real-time IDS / IPS due to performance limitations. Various techniques and approaches have been proposed in the literature to improve real-time smart contract security in the DeFi ecosystem. One approach is the use of rule-based methods, which rely on predefined rules and patterns to detect and prevent smart contract vulnerabilities in real-time. Related work explored a rule-based approach to detect and prevent price oracle manipulation attacks~\cite{wu2021defiranger} as well as machine learning to detect reentrancy attacks~\cite{eshghie2021dynamic}.

It is worth noting that while the above-mentioned methods are effective in identifying specific vulnerabilities, they may not cover all possible types of vulnerabilities. Our proposed work aims to detect any anomaly in transaction trace in real-time while being protocol-agnostic.

\subsection{Embedding Techniques In \NLP}

\begin{description}
\item[ELMo:] ELMo~\cite{peters2018deep} is a deep contextualized model that represents characteristics of word use (e.g., syntax and semantics) across linguistic contexts and captures context-dependent aspects of word meaning. ELMo takes the entire sentence as the input of a bidirectional LSTM (biLSTM) model, thus effectively encoding the contextualized sentence information.
\item[BERT:] By following ELMo, Devlin~\etal~\cite{devlin2018bert} propose a deep pre-trained embedding model called BERT, which applies bidirectional training of Transformer with an attention mechanism that learns contextual dependency between words. Moreover, BERT can be fine-tuned for a wide variety of NLP tasks by adding just one output layer to the core model.
\end{description}


\section{Conclusion}\label{sec:conclusion}
In this work, we introduced \name, an innovative transaction anomaly ranking tool for Ethereum-based blockchains. By analyzing a rich dataset spanning four years, \name demonstrated its ability to accurately identify attack transactions amidst a highly imbalanced dataset of benign and adversarial transactions. Our results indicate that \name effectively detects abnormal transactions, ranking 49 out of 124 attacks among the top-3 most abnormal transactions interacting with their corresponding victim contracts.

\name showcases its real-time capabilities with an average batch throughput of $2,284 \pm 289$ transactions per second, making it a viable real-time intrusion detection system for blockchain networks such as Ethereum. The proposed system can trigger smart contract pause mechanisms in response to malicious blockchain transactions, thus preventing attacks.

Our research contributes to the field of blockchain transaction analysis by being the first to employ unsupervised/self-supervised learning for anomaly detection of transactions. Additionally, we constructed a large language model specifically designed for this task, incorporating custom data encoding and domain-specific tokenization techniques. This work lays the foundation for further exploration of real-time, learning-based security analysis tools for blockchain networks.

\bibliography{references}{}
\bibliographystyle{IEEEtran}

\appendices

\section{$F_{beta}$ and $F_10$}\label{app:f10}
$F_\beta$-score is a generalization of the $F_1$-score, which ``measures the effectiveness of retrieval regarding a user who attaches $\beta$ times as much importance to recall as precision''~\cite{rijsbergen1979information}. Figure~\ref{fig:f10} visualizes the best $F_10$ score that \name can achieve and the corresponding precision and recall, which attaches~$10$ times as much importance to recall as precision.

\begin{figure}[tb]
    \centering
    \includegraphics[width=\columnwidth]{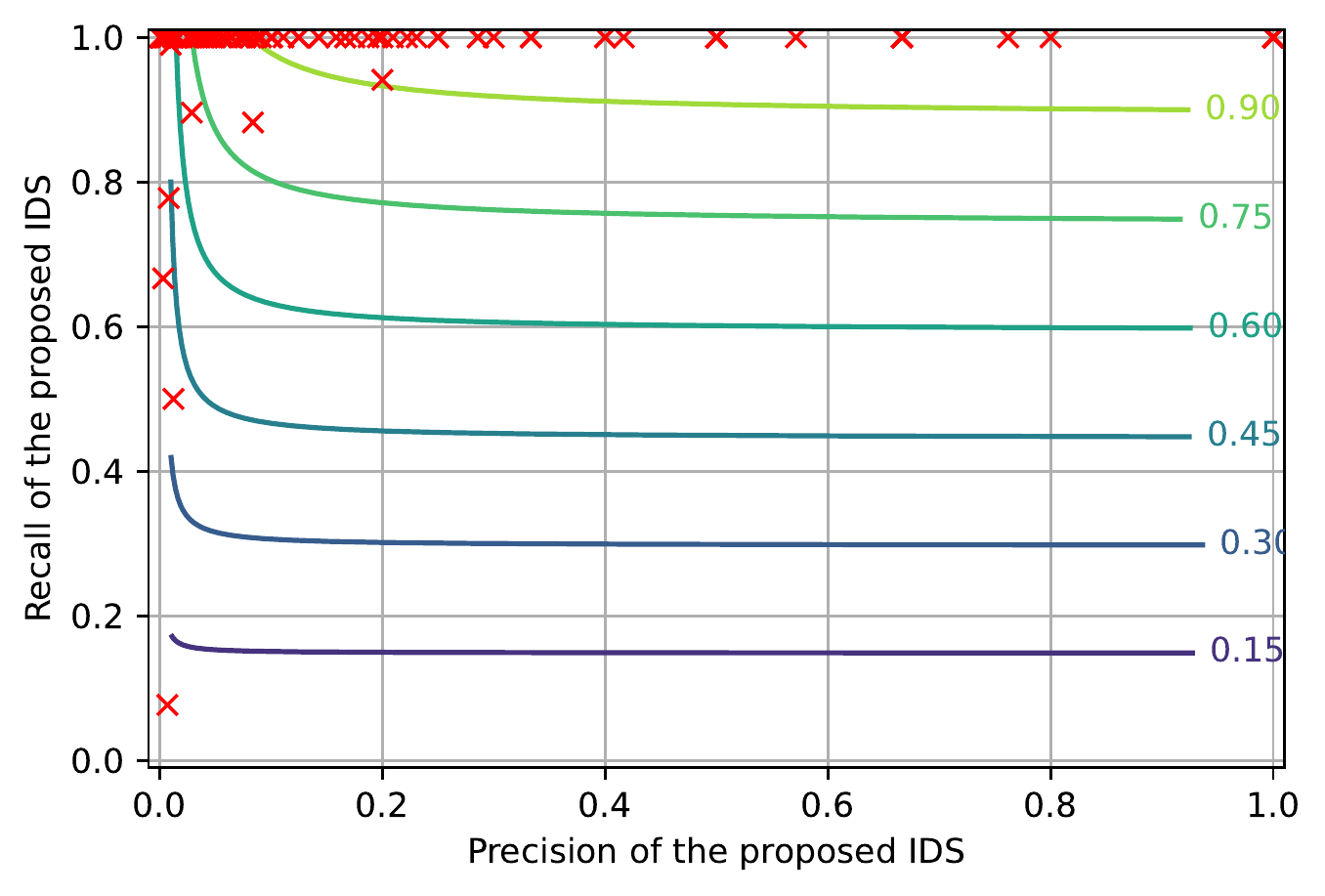}
    \caption{
        Identical to Fig.~\ref{fig:f1}, except with the $F_1$-score replaced with the $F_{10}$ score ($F_\beta$ score with $\beta = 10$).
        Instead of weighing the precision and recall of the IDS equally like the $F_1$ score, the $F_{10}$ score attaches $10$ times as much importance to recall as precision, reflecting the fact that an undetected attack costs significantly more than a false alarm for \DeFi applications.}
    \label{fig:f10}
\end{figure}

\section{Intrusion Detection Capability ($\cid$)}\label{app:cid}
Traditional metrics such as false alarm rate, detection rate, and F-score may not provide a complete picture when the dataset is imbalanced and the cost of false negatives is high~\cite{cardenas2006framework}. In such cases, a high detection rate is desirable as it indicates that the \IDS is better at detecting intrusions, even if it comes at the cost of a higher false alarm rate, lower F1, and lower precision. To address this, researchers have proposed advanced metrics such as $\cid$~\cite{gu2006measuring}, which considers the operating environment and costs associated with false alarms and missed intrusions. In this paper, we propose to apply CID to evaluate the performance of our \DeFi \IDS, considering the imbalanced nature of the dataset and the high cost of false negatives. This aims to provide a more comprehensive evaluation and improve decision-making for the DeFi \IDS's deployment and maintenance.

On a high-level, the $\cid$ is defined as the ratio of the mutual information between the \IDS input and output to the entropy of the input, where $I$ and $H$ respectively denote the mutual information and the entropy.
\[
C_{\text{ID}} = \frac{I(X; Y)}{H(X)}
\]

The entropy $H(X)$ of a random variable $X \in \mathcal X$ is \[
    H(X) = -\sum_{x \in \mathcal X} P_X (x) \log P_X (x)
\] where $P_X$ denotes the distribution of $X$.
Intuitively, the entropy $H(X)$ quantifies the uncertainty in $X$. The mutual information $I(X; Y)$ between discrete random variable $X \in \mathcal X$ and $Y \in \mathcal Y$ is given by \[
    I(X; Y) = \sum_{x \in \mathcal X} \sum_{y \in \mathcal Y} P_{X, Y} (x, y) \log \frac{P_{X, Y} (x, y)}{P_X (x) P_Y (y)}
\] where $P_{X, Y}$ denotes the joint distribution of $X$ and $Y$, and $P_X$ and $P_Y$ denote the marginal distribution of $X$ and $Y$ respectively.
Intuitively, the mutual information $I(X; Y)$ quantifies the amount of information about $Y$ that observing $X$ yields.
The Intrusion Detection Capability $\cid$, being the ration between the mutual information and the entropy, thus quantifies the proportion of uncertainty in a transaction being abnormal that is captured by the \IDS.

Although the $\cid$ accounts for attack distributions by normalizing the mutual information between attacks and alerts with the entropy of attacks, it fails to incorporate the cost of undetected attacks and false alarms.
More generally, we can associate a cost $\gamma_{x, y}$ with each pair of outcomes $(x, y) \in \mathcal X \times \mathcal Y$. 
The \emph{cost-aware} mutual information between $X$ and $Y$ is then \[
    I_\gamma (X; Y) = \frac1\gamma \sum_{x \in \mathcal X} \sum_{y \in \mathcal Y} \gamma_{x, y} P_{X, Y} (x, y) \log \frac{P_{X, Y} (x, y)}{P_X (x) P_Y (y)}
\]

\[
    \gamma = \sum_{x \in \mathcal X} \sum_{y \in \mathcal Y} \gamma_{x, y} 
\] is the total cost of all outcomes.
Multiplying each term by the cost associated with the outcome and dividing by the total cost biases the mutual information towards outcomes with high costs.
With the entropy $H(X)$ of $X$ defined as above, the $\emph{cost-aware}$ Intrusion Detection Capability $\cid^{(\gamma)}$ is given by\[
    \cid^{(\gamma)} = \frac{I_\gamma (X; Y)}{H(X)}
\]

The advantage of the cost-aware $\cid$ is that it only requires cost estimates that are easy to calculate. For an IDS, both X (if a pending transaction is an attack) and Y (if the IDS sends an alert for a pending transaction) are binary variables. The costs we need to estimate include:

\begin{enumerate}
\item X and Y are both positive: The IDS alerts the DeFi protocol operator about a potential attack. After verifying it's an attack, the operator stops it. Assuming no cost to prevent attacks, the cost of this outcome is the cost of the operator inspecting a transaction.
\item X is negative, Y is positive (false positive): The IDS alerts the operator about a possible attack, but it's a benign transaction. The operator lets it proceed. The cost of this outcome is the cost of the operator inspecting a transaction.
\item X is positive, Y is negative (false negative): An attack succeeds without the operator noticing, causing a loss to the protocol. Our data shows that each attack costs around ten million USD on average.
\item X and Y are both negative: The transaction goes through with no cost.
\end{enumerate}

To estimate the cost of a \DeFi protocol operator inspecting a transaction, we use the statistics from a recruitment firm that \DeFi security auditors can earn~$400,000$ USD per year in 2022.
We estimate that inspecting a false positive transaction will take one hour or more, which is approximately~$204$ USD. It should be noted that \DeFi protocol operators can choose to activate defense mechanisms such as emergency pause. These defenses may have an impact on the user experience, resulting in reputation damage and an implicit cost. We ignore the reputation damage in this paper. We estimate the distribution $P_X$, $P_Y$, and $P_{X, Y}$ using the frequency estimator.

\begin{figure}[tb]
    \centering
    \includegraphics[width=\columnwidth]{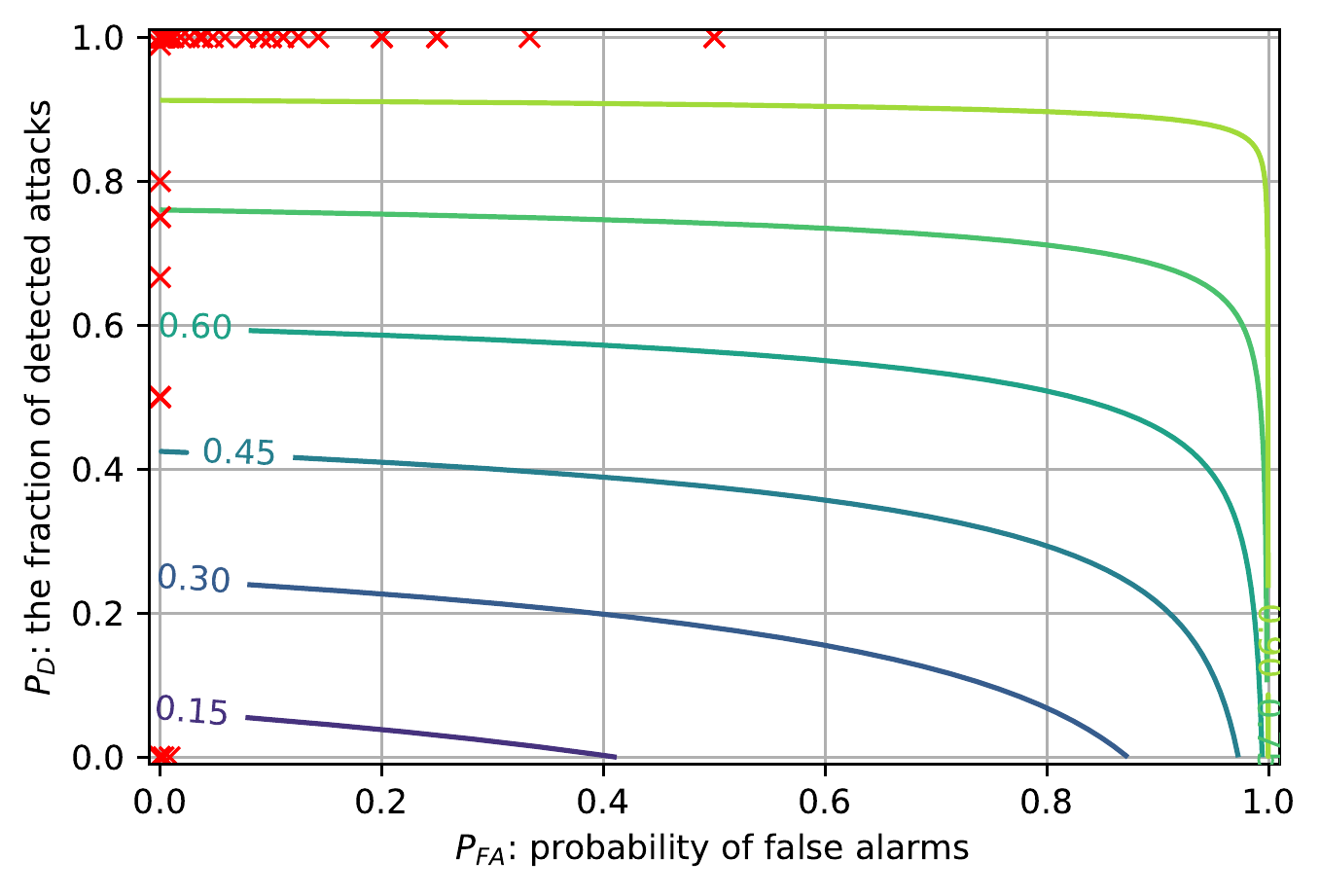}
    \caption{
        The Intrusion Detection Capability ($\cid$) of the proposed IDS.
        The x-axis corresponds to $P_{FA}$, the probability of the IDS raising false alarms (FA), while the y-axis corresponds to $P_D$, the fraction of attacks detected the IDS.
        We also plotted the contour lines of $\cid$ as a bivariate function of $P_{FA}$ and $P_D$.
        For example, given a $C_{ID}$ of $0.3$, we find that if the false positive rate is 0.4, then the probability of a true positive is 0.24.
    }
    \label{fig:cid}
\end{figure}

Figure~\ref{fig:cid} shows the distribution of $\cid$ for~$124$ attacks in relation to the probability of false alarms ($P_{FA}$) and the fraction of detected attacks ($P_D$). The figure allows for an easy comparison of the performance of the \IDS across different smart contracts. By using an example of a $C_{ID}$ of $0.3$, we can see that if the false positive rate is~$0.4$, the probability of a true positive is~$0.24$. It is important to note that, despite a high probability of false alarms,~$117$ out of the~$124$ attacks achieved a $\cid$ score of more than~$0.9$. This is because the cost of a false negative is much higher than the cost of a false positive, which results in a higher weighting of $P_D$.

\section{Baseline - doc2vec}\label{app:baseline}
To compare \name to a naive baseline, as a proof of concept, we prototyped a transaction ranker that estimates the likelihood of transactions using doc2vec and Gaussian mixture. The prototype ranks~$8$ as the least likely (most abnormal) transaction that interacted with their victim contracts,~$11$ as the second least likely, and~$5$ as the third least likely. Our results show that \name identifies abnormal transactions by ranking $49$ out of $124$ attacks among the top-3 most abnormal transactions interacting with their victim contracts.

\subsubsection{Word2Vec} A prominent embedding approach to learn the distributed representation of words is Word2Vec. Mikolov \etal~\cite{mikolov2013efficient,mikolov2013distributed} propose two models: the continuous bag-of-words (CBOW) model that predicts the center word given its surrounding context and the skip-gram (SG) model that predicts the surrounding context given a center word.

\subsubsection{Doc2Vec}
Motivated by Word2Vec, Le \etal~\cite{le2014distributed} propose Doc2Vec that represents input documents as dense vectors.
Doc2Vec also has two models: the distributed memory model of paragraph vectors (PV-DM) and distributed bag of words version of paragraph vector (PV-DBOW).
PV-DM is technically similar to CBOW while PV-DBOW is similar to SG. 
It is used to learn words' representations of arbitrary length sequences.
Concretely: Given a set of documents $\mathbb{D} = \{ d_{1}, d_{2}, d_{3},...\} $ and a sentense $s(d_{i}) =  \{ w_{1}, w_{2}, w_{3},...\}$ composed of words from document $d_{i}$, it learns the embeddings $\vec{d_{i}} \in \mathbb{R}^\delta$ and $\vec{w_{i}} \in \mathbb{R}^\delta$. This is done by considering a word $w_{i} \in s(d_{i})$ and maximizing $\sum_{j=1}^{l} \log Pr(w_{i} | d_{i})$ where the probability of $w_i$ occuring in the context $d_i$ is $Pr(w_{i} | d_{i}) = \frac{\exp \vec{d_{i}} . \vec{w_{i}}}{\sum_{w \in \mathbb{V}}^{} \exp \vec{d} . \vec{w}}$. The key computational challenge to the approach is to compute the denominator $\sum_{w \in \mathbb{V}}^{} \exp \vec{d} . \vec{w}$, as the vocabulary size $|\mathbb V|$ can grow drastically as the corpus size grows.
\cite{le2014distributed} proposes two solutions: \begin{itemize}
    \item Hierarchical softmax.
        Given a binary tree with each leaf associated with a word in the vocabulary, a hierarchical softmax classifier assigns a vector to each node of the tree.
        Let $v_n$ denote the vector assigned to node $n$.
        In the following we omit the context for brevity.
        All probabilities should be understood as conditional on the context.
        Given a word $w$ and a path $n_1, \ldots, n_k$ from the root of the tree to the leaf associated with the word $w$, the log-likelihood of the word occuring is given by \begin{align*}
            \log \Pr[w]
            & = \log \Pr[n_1, \ldots, n_k | w] = \log \prod_{j = 1}^k \Pr[n_j | w] \\
            & = \sum_{j = 1}^k \log \Pr[n_j | w] = \sum_{j = 1}^k \log \sigma(v_{n_j}^T w)
        \end{align*}
        where $\sigma$ is the sigmoid function given by \[
            \sigma (x) = \frac{\exp x}{\exp x + 1} \in (0, 1)
        \]
    \item Negative sampling.
        Negative sampling reduces the computation cost of the denominator $\sum_{w \in \mathbb{V}}^{} \exp \vec{d} . \vec{w}$ by only summing over a random sample of the vocabulary $\mathbb V$.
        Mathematically, \[
            Pr(w_{i} | d_{i}) = \frac{\exp \vec{d_{i}} . \vec{w_{i}}}{\sum_{w \in \mathbb{V'}}^{} \exp \vec{d} . \vec{w}}
        \] where $V' \subset V$ is chosen randomly.
        Clearly, the efficiency of the solution depends on how to draw samples from the vocabulary.
        \cite{le2014distributed} proposes to draw samples from the vocabulary following the rule below: \[
            P(w) = 1 - \sqrt\frac{t}{f(w)}
        \] where $f(w)$ is the frequency of word $w$ in a training corpus, and $t$ is a hyperparameter.
        Intuitively, the sampling rule increases the likelihood of rare words being chosen compared to uniform sampling, which makes sense as rare words are more likely to be informative than frequent ones.
\end{itemize}

We experimented with both approaches and observed no significant performance difference in our settings. The baseline result reported in this paper was achieved using negative sampling, which provided marginally better results.

\subsubsection{Doc2Vec and gaussian mixture}

Mathematically, the log-likelihood of a trace with doc2vec embedding $v \in \mathbb{R}^d$ under a Gaussian mixture model with parameters $\pi_1, \ldots, \pi_C \in \mathbb R$, $\mu_1, \ldots, \mu_C \in \mathbb{R}^d$, and $\Sigma_1, \Sigma_C \in \mathbb{R}^{d \times d}$ is given by
\[\log p(v|\pi, \mu, \Sigma) = \log \sum_{c = 1}^C \pi_C \phi_{\mu_c, \Sigma_c} (v)\]
where $\phi_{\mu, \Sigma}$ is the multidimensional Gaussian probability density function with mean $\mu$ and covariance $\Sigma$, given by \[
    \phi_{\mu, \Sigma} (x) = (2 \pi)^{-d / 2} (\det \Sigma)^\frac12 \exp \left(-\frac12 (x - \mu)^T \Sigma (x - \mu)\right)
\] where $\det \Sigma$ is the determinant of the covariance matrix $\Sigma$.

Gaussian mixture models can be interpreted as probabilistic clustering of vector embeddings, with the parameter $C$ interpreted as the number of clusters, the parameters $\pi_1, \ldots, \pi_C$ the probability of the vector embedding belonging to each cluster, and the parameters $\mu_1, \ldots, \mu_C$ the centroid of each cluster.
The parameters $\pi$, $\mu$, and $\Sigma$ are estimated by maximizing the log-likelihood of historical transactions: \[
  \max_{\pi, \mu, \Sigma} \sum_{v} \log p(v|\pi, \mu, \Sigma)
\]
which is accomplished by an Expectation-Maximization (EM) algorithm \cite{dempster1977maximum}.

The EM algorithm alternates between an E (Expectation) and M (Maximization) step until some convergence criterion is met.
The E step assigns each observation to the cluster where its likelihood is maximized.
Mathematically, the cluster $c_x$ assigned to an observation $x$ is given by \[
    c_x = \argmax_{c = 1, \ldots, C} \log \pi_c + \log \phi_{\mu_c, \Sigma_c} (x)
\]
The M step that follows then update the parameters $\pi_1, \ldots, \pi_C$, $\mu_1, \ldots, \mu_C$, and $\Sigma_1, \ldots, \Sigma_C$ by maximizing the likelihood of all observations.
The likelihood of each cluster $\pi_1, \ldots, \pi_C$ is simply given by the proportion of observations that are assigned in the cluster.
The mean and covariance of each cluster are given by maximizing the log-likelihood: \begin{align*}
    \log \phi_{\mu_c, \Sigma_c} (\{x: c_x = c\})
    & = \log \prod_{x: c_x = c} \phi_{\mu_c, \Sigma_c} (x) \\
    & = \sum_{x: c_x = c} \log \phi_{\mu_c, \Sigma_c} (x)
\end{align*}

In our experiments, we initialize the clusters with the K-Means algorithm, and leave experimenting other initialization methods for future work.
More specifically, we first cluster the embeddings of traces in our training corpus with K-Means, and then set the cluster means $\mu_1, \ldots, \mu_C$ to the centroids of the clusters.
The likelihood of each cluster $\pi_1, \ldots, \pi_C$ is set to the proportion of observations that are assigned to the cluster, and the covariance of each cluster is set to the sample covariance of each cluster.
The objective of the K-Means algorithm is \[
    \argmin_S \sum_{i = 1}^K \sum_{x \in S_i} |x - \mu_i|^2
\] where $S = (S_1, \ldots, S_K)$ is a tuple of $K$ disjoint sets (clusters) such that $\cup_{i = 1}^K S_i = \{x_j\}_{j = 1}^n$

Multiple parameterization schemes for the covariance matrix exist.
The one most expressive, yet most prone to overfit, is to use a full covariance matrix.
The least expressive one is to reduce the covariance matrix to a single positive number.
An approach that attempts to strike a balance between these extremes is to constrain the covariance matrix to be diagonal, i.e. \[
    \Sigma =
    \begin{bmatrix}
        \Sigma_{1} & & \\
        & \ddots & \\
        & & \Sigma_{d}
    \end{bmatrix}
\]
To avoid overfitting while maintaining some expressive power, we constrain the covariances $\Sigma_1, \ldots, \Sigma_C$ to be diagonal.

The parameter $C$ is chosen by minimizing the Bayesian information criteria (BIC) \cite{neath2012bayesian} \[
  \min_C -2 \sum_{v} \log p(v|\pi, \mu, \Sigma) + d(C) \log N
\] where $N$ is the number of historical transactions and $d(C)$ is the number of parameters of a Gaussian mixture model, determined by the parameter $C$. Intuitively, the BIC consists of two parts, a log-likelihood term and a regularization term \[
    \sum_{v} \log p(v|\pi, \mu, \Sigma), \quad d(C) \log N
\]  which penalizes mixtures with large number of parameters, e.g. mixtures with full covariance matrices.
The regularization term grows as the number of observations grows to keep up with the scale of the log-likelihood term.

We find in preliminary experiments that diagonal covariance matrices generally maximize the BIC, likely due to a balance between their complexity and expressiveness.
In our case, where all covariance matrices are constrained to be diagonal, the number of parameters $d(C)$ is given by \[
    d(C) = C + C \cdot d + C \cdot d = C \cdot (2 d + 1)
\]
Without any constraint on the covariance matrices, the number of parameters $d(C)$ is given by \[
    d(C) = C + C \cdot d + C \cdot d^2 = C \cdot (d^2 + d + 1)
\]
When all covariance matrices are reduced to a single positive number, the number of parameters $d(C)$ is given by \[
    d(C) = C + C \cdot d + C = C \cdot (d + 2)
\]

\section{Normalization in Transformer Encoder}
We find empirically that normalization is crucial to training the transformer encoder, particularly given our unique positional encoding scheme. As the sum of many embeddings, embeddings produced by our tree encoding can easily reach numerical scales that destabilize training. We find Layer Normalization (LayerNorm) \cite{ba2016layer} particularly helpful in this case. Given a $d$-dimensional vector $v$, LayerNorm normalizes the vector as $\hat v = \frac{v - \mu}\sigma$, where $\mu$ is the mean of all components of $v$, given by $\mu = \frac1d \sum_{i = 1}^d v_i$ and $\sigma$ is the standard deviation of all components of $v$, given by $\sigma = \sqrt{\frac1d \sum_{i = 1}^d (v_i - \mu)^2}$. Clearly, the normalized vector $\hat v$ satisfies the conditions $\hat\mu = \frac1d \sum_{i = 1}^d \hat{v}_i = 0$ and $\hat\sigma = \sqrt{\frac1d \sum_{i = 1}^d (\hat{v}_i - \mu)^2} = 1$, which contributes to more stable training of the transformer encoder.

\section{Second and Third Most Abnormal Transactions}\label{app:top_2_3}
We present the results of our analysis on the transactions that \name identified as the second and third most abnormal. Figure \ref{tab:top2} showcases the 20 attacks ranked as the second most abnormal transaction, while Figure \ref{tab:top3} highlights the 7 attacks that were ranked as the third most abnormal transaction. These tables provide insights into the victim names, victim contracts, application categories, and damages (in USD) associated with each attack, further demonstrating the effectiveness of \name in detecting various types of malicious activities on the Ethereum blockchain.

\begin{figure}[tb]
    \centering
    \resizebox{\columnwidth}{!}{%
    \begin{tabular}{lllll}
        \toprule
           Victim Name &                            Victim Contract &             \makecell{Application\\Categories} & \makecell{Damage\\(in USD))} \\
        \midrule
         SorbetFinance & \abbrEtherscanAddress{0x14e6d67f824c3a7b4329d3228807f8654294e4bd} &                             Others &   27,000,000 \\
        InverseFinance & \abbrEtherscanAddress{0x39b1df026010b5aea781f90542ee19e900f2db15} &                            Lending &   15,600,000 \\
           WarpFinance & \abbrEtherscanAddress{0xae465fd39b519602ee28f062037f7b9c41fdc8cf} &                            Lending &    7,800,000 \\
              DAOMaker & \abbrEtherscanAddress{0xa43b89d5e7951d410585360f6808133e8b919289} &                             Others &    4,000,000 \\
              DAOMaker & \abbrEtherscanAddress{0x2fd602ed1f8cb6deaba9bedd560ffe772eb85940} &                             Others &    4,000,000 \\
               GemSwap & \abbrEtherscanAddress{0x775541df8bd9a39ae9b15556628e6abe21074dab} &                                DEX &    1,300,000 \\
               GemSwap & \abbrEtherscanAddress{0xd361a3cffe2dd66a5beb82f894839cd3e56611e1} &                                DEX &    1,300,000 \\
               GemSwap & \abbrEtherscanAddress{0x42653f548139de55104204fb3a828715f5bceab6} &                                DEX &    1,300,000 \\
               GemSwap & \abbrEtherscanAddress{0xf3d161c4bb9a67921730fb7a0faa696996b20e12} &                                DEX &    1,300,000 \\
               GemSwap & \abbrEtherscanAddress{0xa41638f9eeeecbdf0cb5e5ddc959795fc096bdf6} &                                DEX &    1,300,000 \\
               GemSwap & \abbrEtherscanAddress{0xefcb3057fe5e65b04a7ea0f8f02d41c541ec5b8f} &                                DEX &    1,300,000 \\
               GemSwap & \abbrEtherscanAddress{0x748fb6bce7606cf439957f48e821d270fcd0190d} &                                DEX &    1,300,000 \\
               GemSwap & \abbrEtherscanAddress{0x87704c841cc0a4d8f4fc690b8e7679c7dd78fb90} &                                DEX &    1,300,000 \\
               GemSwap & \abbrEtherscanAddress{0x8cc238391ac1c576ce43072eddaf082dc4c00a19} &                                DEX &    1,300,000 \\
               GemSwap & \abbrEtherscanAddress{0xe777b006905e0f7f626f3af8824d3ae870e5b02d} &                                DEX &    1,300,000 \\
             BasketDAO & \abbrEtherscanAddress{0x4622aff8e521a444c9301da0efd05f6b482221b8} &                                DAO &    1,200,000 \\
            Li.Finance & \abbrEtherscanAddress{0x5a9fd7c39a6c488e715437d7b1f3c823d5596ed1} &                     DEX aggregator &      600,000 \\
          BuildFinance & \abbrEtherscanAddress{0x3157439c84260541003001129c42fb6aba57e758} &                                DAO &      470,000 \\
           SashimiSwap & \abbrEtherscanAddress{0xe4fe6a45f354e845f954cddee6084603cedb9410} &                                DEX &      200,000 \\
          Formation.Fi & \abbrEtherscanAddress{0xcb6afdc84e8949ddf49ab00b5b351a5b0f65a723} &                      Yield farming &      100,000 \\
        \bottomrule
        \end{tabular}
        }
        \caption{The $20$ attacks ranked by \name \IDS as the \emph{second} most abnormal transaction that interacted with the respective victim contract.}
        \label{tab:top2}
\end{figure}

\begin{figure}[tb]
    \centering
    \resizebox{\columnwidth}{!}{%
    \begin{tabular}{lllll}
        \toprule
           Victim Name &                            Victim Contract &  \makecell{Application\\Categories} &  \makecell{Damage\\(in USD))} \\
        \midrule
        IndexedFinance & \abbrEtherscanAddress{0x5bd628141c62a901e0a83e630ce5fafa95bbdee4} &                 Others &   16,000,000 \\
             ValueDeFi & \abbrEtherscanAddress{0xddd7df28b1fb668b77860b473af819b03db61101} &          Yield farming &    7,200,000 \\
              DAOMaker & \abbrEtherscanAddress{0xdd571023d95ff6ce5716bf112ccb752e86212167} &                 Others &    4,000,000 \\
              DAOMaker & \abbrEtherscanAddress{0x6e70c88be1d5c2a4c0c8205764d01abe6a3d2e22} &                 Others &    4,000,000 \\
               GemSwap & \abbrEtherscanAddress{0x8cc757bc682e718d2f0264c1fcd3269eccf8214e} &                    DEX &    1,300,000 \\
                Vether & \abbrEtherscanAddress{0x75572098dc462f976127f59f8c97dfa291f81d8b} &                 Others &      900,000 \\
             Chainswap & \abbrEtherscanAddress{0xc5185d2c68aaa7c5f0921948f8135d01510d647f} &     Cross chain bridge &      800,000 \\
        \bottomrule
    \end{tabular}
    }
    \caption{The $7$ attacks ranked by \name \IDS as the \emph{third} most abnormal transaction that interacted with the respective victim contract.}
    \label{tab:top3}
\end{figure}

\end{document}